%% file: main.tex
\documentclass[twocolumn]{aastex631}
\usepackage[inkscapelatex=false]{svg}
\usepackage{array}
\usepackage{booktabs}
\usepackage{threeparttable}
\usepackage{tabularx}
\usepackage{caption}
\usepackage{ragged2e}
\DeclareCaptionFont{custom}{\footnotesize}
\usepackage{amsmath}
\usepackage{svg}
\usepackage{gensymb}
\usepackage{natbib}
\usepackage{appendix}
\newcommand{\comment}[1]{}
\usepackage{etoolbox}

\usepackage{placeins}
\usepackage{hyperref}
\begin{document}

\title{IBEX Observations of Elastic Scattering of Interstellar Helium by Solar Wind Particles}
\author{H. Islam}
\affiliation{Physics Department, Space Science Center,\
University of New Hampshire,\
Durham, NH 03824, USA }

\author{N. Schwadron}
\affiliation{Physics Department, Space Science Center,\
University of New Hampshire,\
Durham, NH 03824, USA }
\author{E. Möbius}
\affiliation{Physics Department, Space Science Center,\
University of New Hampshire,\
Durham, NH 03824, USA }

\author{F. Rahmanifard}
\affiliation{Physics Department, Space Science Center,\
University of New Hampshire,\
Durham, NH 03824, USA }

\author{J. M. Sok\'{o}\l}
\affil{Southwest Research Institute,\
San Antonio, TX 78228, USA}

\author{A. Galli}
\affiliation{Physics Institute,\
University of Bern,\
Bern, 3012, Switzerland }

\author{D. J. McComas}
\affil{Department of Astrophysical Sciences,\
Princeton University,\
Princeton, NJ 08544, USA}

\author{P. Wurz}
\affil{Physics Institute,\
University of Bern,\
Bern, 3012, Switzerland }

\author{S. A. Fuselier}
\affil{Southwest Research Institute,\
San Antonio, TX 78228, USA}
\affil{University of Texas at San Antonio,\
San Antonio, TX 78228, USA}

\author{K. Fairchild}
\affiliation{Physics Department, Space Science Center,\
University of New Hampshire,\
Durham, NH 03824, USA }

\author{D. Heirtzler}
\affiliation{Physics Department, Space Science Center,\
University of New Hampshire,\
Durham, NH 03824, USA }

\begin{abstract}
The IBEX-Lo instrument on the Interstellar Boundary Explorer (IBEX) mission observes primary and secondary interstellar helium in its 4 lowest energy steps. Observations of these helium populations have been systematically analyzed and compared to simulations using the analytic full integration of neutrals model (aFINM). A systematic difference is observed between the simulations and observations of secondary He during solar cycle (SC) 24.  We show that elastic scattering of primary helium by solar wind protons, which redistributes atoms from the core of the flux distribution, provides an  explanation of the observed divergence from simulations. We verify that elastic scattering forms a halo in the wings of the primary He distribution in the spin-angle direction. Correcting the simulation for the effects of elastic scattering requires an increase of the estimated density of primary helium compared to previous estimates by Ulysses/GAS. Thus, based on our analysis of IBEX observations and $\chi ^2$ minimization of simulation data that include the effects of elastic scattering, any estimation of neutral interstellar helium density at 1 AU by direct detection of the peak flux of neutral helium needs to be adjusted by $\sim$ 10\%.
\end{abstract}
\input{sections/1intro}
\input{sections/2data}

\input{sections/3temporal_variation}
\input{sections/4scattering}

\input{sections/5density}

\input{sections/7conclusion}

\vspace{5mm}
\bibliography{sources}{}
\bibliographystyle{aasjournal}


\end{document}

%% file: sections/1intro.tex
\section{Introduction} \label{sec:intro}

The Sun produces the supersonic solar wind (SW), which flows out radially and inflates the volume of our heliosphere. The distant boundaries of our heliosphere are created through the interaction of the SW with the dilute partially ionized gas of the very local interstellar medium (VLISM),  consisting of neutral atoms, such as H, He, N, O, and Ne, and a charged plasma part consisting of electrons, protons, and heavier ions \citep{2010ApJ...719.1984F}. The neutral atoms of the interstellar medium travel relatively unimpeded through the heliosphere. The particle densities are so low in the VLISM and in the solar wind that typical mean collisional free paths are typically 10's to 100's of AU.   

The Interstellar Boundary Explorer (IBEX) is a small NASA explorer mission that has provided us a decade long observation of Interstellar Neutral (ISNs) and Energetic Neutral Atoms (ENAs) in the energy range 15 eV to 6 keV \citep{2009SSRv..146...11M}. There are  two ENA cameras on IBEX: IBEX-Lo which is senstive to neutral atoms in the energy ranges between 10 eV to 2 keV \citep{2009SSRv..146..117F} and IBEX-Hi in the energy range between 0.38 keV to 6 keV \citep{2009SSRv..146...75F}. Since its launch and commissioning,  IBEX has observed ISNs from the pristine interstellar medium and from the interaction zone of the heliospheric boundary and the surrounding VLISM. 

ISNs are an efficient tool to analyse the characteristics of the interstellar medium and the interaction at the boundary. IBEX fundamentally observes three types of neutral atoms (a) pristine ISNs like Hydrogen, Helium, Oxygen and  Neon, (b) secondary populations of ISNs, which are created through charge exchange between primary ISNs  and plasma in the outer heliosheath, and (c) energetic neutral atoms  created from solar wind ions that undergo charge exchange in the inner heliosheath (to produce the globally distributed flux of ENAs),  and form the IBEX ribbon \citep{2009Sci...326..959M,2009Sci...326..966S,2014RvGeo..52..118M}. IBEX observes  hydrogen in ESA step 1 and 2 (15, and 30 eV) and helium in ESA steps 1-4 (15, 30, 55, 110 eV). IBEX-Lo also observed Oxygen in ESA steps 5 and 6. \\

ISNs are  unaffected by magnetic and electric fields. Inside the heliosphere, ISNs undergo charge exchange and photo-ionization by solar extreme ultraviolet (EUV) radiation and thereby create pick-up ions (PUI) that are carried away by the SW. The newly made PUIs mass load and slow the SW plasma as it travels out to the Termination Shock (TS). Beyond the TS, the inner heliosheath (IHS) is populated by accelerated PUIs as well as decelerated and deflected SW. The IHS plasma is separated from the interstellar medium by a tangential discontinuity called the HelioPause (HP). 

Among all species, helium is the second most abundant element in the interstellar medium after hydrogen (He/H $\sim$ 10) and is the least prone to ionization due to its high ionization potential and very low cross-section for charge exchange with the SW ions. Consequently, helium has the highest intensity at 1 AU and serves as an exceptional tool for studying the physical properties of interstellar medium. The first detection of neutral interstellar helium began with sounding rockets \citep{1974ApJ...193..471W}, followed by satellites \citep{1973ApJ...183L..87P,1974ApJ...188L..71P}, both methods relying on absorption and re-emission of solar EUV by helium. The discovery of helium pickup ions ($\text{He}^+$) in the SW by \citet{1985Natur.318..426M} introduced a novel method for analyzing neutral helium via the direct detection of $\text{He}^+$.

The first direct detection of neutral interstellar helium was made by the GAS experiment onboard Ulysses \citep{1992A&AS...92..333W}. The analysis of the GAS/Ulysses observation set the standard for interstellar helium parameters \citep{1993AdSpR..13f.121W,2004A&A...426..835W}. The density was determined to be $0.015 \pm 0.0028$ $\text{cm}^{-3}$, which was obtained as a best fit across two different observation seasons where the photoionization rate ranged from 0.6 to 1.6 $\times 10^{-7} \text{s}^{-1}$. Charge exchange and electron impact ionization were disregarded due to their negligible contributions. This density estimation aligns closely with the estimation of \citet{2004AdSpR..34...53G},  $0.0154 \pm 0.0015$ cm$^{-3}$, based on measurements of interstellar He$^{++}$ PUIs.
The interstellar plasma flow outside  the HP contains an abundance of 
He$^+$ ions \citep{2003ApJ...594..844F} that charge exchange with neutral helium atoms and thereby create a secondary  population of neutral helium, 
 dubbed ``the warm breeze"  \citep{2014ApJS..213...29K,2016ApJS..223...25K}.  The warm breeze is warmer, slower and also deflected from the interstellar primary flow. 
In the progression of each ISN season, IBEX-Lo first observes the secondary He in early December to Mid January, followed by the primary population from early February to mid-March. Then the signatures of hydrogen are  prominent until end of May \citep{2012ApJS..198...14S,2012ApJS..198...14S}. 

The Sun goes through an approximately 11-year cycle during which its activity varies between relatively low levels (solar minimum) of activity to higher levels (solar maximum). During solar maxima, both solar wind fluxes and solar radiation fluxes increase. EUV radiation, which is tied to the photoionization rate, nearly doubles during solar maxima. ISN He within the heliosphere is modulated primarily by photoionization, a pattern that is evident in IBEX-Lo observations throughout Solar Cycle (SC) 24, spanning from 2009 to 2019 \citep{2022ApJS..259...42S,2019ApJ...887..217R}. 

We analyze the temporal variation of the differential flux  of primary and secondary helium. The flux of the primary helium population shows a gradual increase starting from 2015, while the flux of the secondary population appears higher than expected. The temporal variation of secondary helium turns out to be very sensitive to the primary helium contribution. during the solar maximum and then decreases. The  variation of the photoionization rate over the solar cycle can explain partially the observed changes in the primary population \citep{2022ApJS..259...42S}, but does not account for the changes in the secondary population. As a potential explanation, we investigate the effects associated with elastic scattering of primary helium by the solar wind \citep{1986P&SS...34..387G}. \\

As detailed by \citet{https://doi.org/10.1002/jgra.50199}, the inclusion of scattering and the redistribution of primary helium  can be interpreted as a loss of counts in the core of the primary helium velocity distribution. Quantitatively, this loss is similar in nature but much weaker than the photoionization loss. A direct consequence of incorporating loss due to elastic scattering is the revision of the interstellar helium density derived from direct neutral He observations, as reported for the first time in this study.\\
In Section \ref{sec:data}, we describe the observations used for this study. Then we outline the model used to compare observations with simulations in Section \ref{sec:model}. A central quantity used throughout this study is a renormalization constant ($A$), which multiplies the amplitude of modeled He fluxes and is determined through $\chi^2$ minimization. Section \ref{sec:temporal} details the temporal variation of the renormalization constants for primary and secondary helium. In order to explain the increase of renormalization factor for secondary helium we discuss elastic scattering of solar wind proton and interstellar helium and also compares theoretical predictions with observations in section \ref{sec:scattering}. In section \ref{sec:density}, we explain why incorporating scattering requires the density estimation of primary helium and which previous studies require revision based on our findings. Finally, in Section \ref{sec:conclusion}, we discuss and conclude our results.

%% file: sections/2data.tex
\section{Data} \label{sec:data} 
 IBEX is a Sun pointing spinner that completes a full rotation approximately every 15 seconds, sweeping its field of view across a roughly fixed swath (roughly centered on a great circle that passes through the ecliptic poles) of the sky. The swath is divided into 60 bins each $6\degree$ wide. IBEX-Lo and IBEX-Hi, the two cameras on IBEX, detect neutral atoms by converting them into ions in the instruments, which are then filtered by energy-per-charge using traditional electrostatic deflection techniques. IBEX-Lo spans the energy range from 0.01 to 2  keV into 8 different  logarithmically spaced energy channels. After a neutral atom passes through the IBEX-Lo collimator it hits the diamond-like conversion surface at a shallow $\sim$15\degree angle of incidence \citep{2009SSRv..146..117F,10.1063/1.1996855}. Neutral atoms with high electron affinity, e.g. hydrogen and oxygen are converted into negative ions with a high probability. Conversely, as a noble gas, He does not produce a stable negative ion. However, it can produce ions by sputtering, predominantly $\text{H}^{-}$, C$^{-}$ and O$^{-}$ ions. The sputtered ions are selected, based on their energy, using the electrostatic analyzer (ESA). The ions that pass through the ESA are then drawn into the time-of-flight (TOF) chamber with an applied Post-Acceleration (PAC) voltage. 
 
 Inside the TOF chamber these particles hit the first carbon foil and release secondary electrons from the surface, which are guided to  the microchannel plate (MCP), creating a ``start-A" signal. Then the ions hit a second carbon foil and the emitted electrons  trigger another ``start-C" signal. Finally, the ion reaches the MCP generating a ``stop-B" signal. 
 
 The MCP has 4 quadrants and the ion is collected in one of these. The position of detection determines the TOF3 delay time along the delay line. If an event has two or three valid time-of-flight (TOF) values, then it is referred to as a double event or triple event, respectively. Triple events that satisfy the condition where the sum of TOF0 and TOF3 equals the sum of TOF1 and TOF2 are known as golden triple events. These golden triple events are characterized by a signal-to-noise ratio (SNR) of 1000 at peak ISN He flow, making them the most reliable events for analysis.

We have used histogram binned (HB) data which are accumulated events in the $6^\circ$ spin bin histogram. Each spin starts at $-3^\circ$ of the North Ecliptic Pole (NEP), which is determined by the Attitude Control System (ACS) onboard IBEX using a star tracker   \citep{2012ApJS..198....9H}. At times when the Moon or the Earth fall into the  the field of view of  the star tracker the attitude information is faulty, resulting in unreliable histogram data. Time tagged direct event (DE) data still can be used by  de-spinning periods when the star tracker is faulty \citep{Fuselier:2009b}.  We have restricted the analyzed data to periods with no star tracker outages.

We have used non despun histogram binned data consisting of golden triple events only. Additionally we remove times when one of the following conditions occur \citep{2022ApJS..261...18G}:
\begin{enumerate}
    \item IBEX is inside Earth's magnetosphere.
    \item Earth or Moon is in the field of view of the IBEX-Lo collimator boresight.
    \item A gain or threshold test  is performed
    \item Counts are larger than 4 in either ESA step 7 or 8 in any of the 64 spin histogram blocks.
    \item The whole orbit is discarded if the same high count as above continues for more than 12 hours, which is approximately 48 histogram blocks.
\end{enumerate}

\begin{table*}[ht]
    \centering
    \caption{Orbits used in this study }
    \begin{tabular*}{\textwidth}{@{\extracolsep{\fill}}ccc@{}}
        \toprule
        \toprule
        \textbf{Year} & \textbf{Primary Orbits } & \textbf{Secondary Orbits } \\
        \midrule
        2013 & 0194b, 0195a, 0195b & 0189a, 0190a, 0191a, 0191b, 0192a, 0192b \\
        2014 & 0235a, 0235b & 0229a, 0230a, 0230b, 0231b, 0232a, 0232b \\
        2015 & 0275a, 0275b & 0269a, 0270b, 0271a, 0271b, 0272a \\
        2016 & 0315a & 0309a, 0309b, 0310a, 0310b, 0311a, 0311b, 0312b, 0313a \\
        2017 & 0355a, 0355b & 0350a, 0350b, 0351a, 0351b, 0352a, 0352b, 0353a \\
        2018 & 0395b, 0396a & 0389b, 0391a, 0391b, 0392a, 0392b, 0393b \\
        2019 & 0435b, 0436a, 0436b & 0430a, 0430b, 0431a, 0431b, 0432a, 0432b, 0433a, 0433b \\
        2020 & 0476a, 0476b & 0470a, 0470b, 0471a, 0472a, 0472b, 0473a, 0473b \\
        \bottomrule
    \end{tabular*}
    \label{tab:orbit}
\end{table*}
The last two conditions are used to determine periods of higher backgrounds.
The time periods free of these conditions, and thus suitable for scientific analysis for each orbit, will henceforth be referred to as ``ISN best-times''.

\begin{figure*}[t]
  \centering
  \begin{minipage}{0.95\textwidth}
    \centering
    \includegraphics[width=\textwidth]{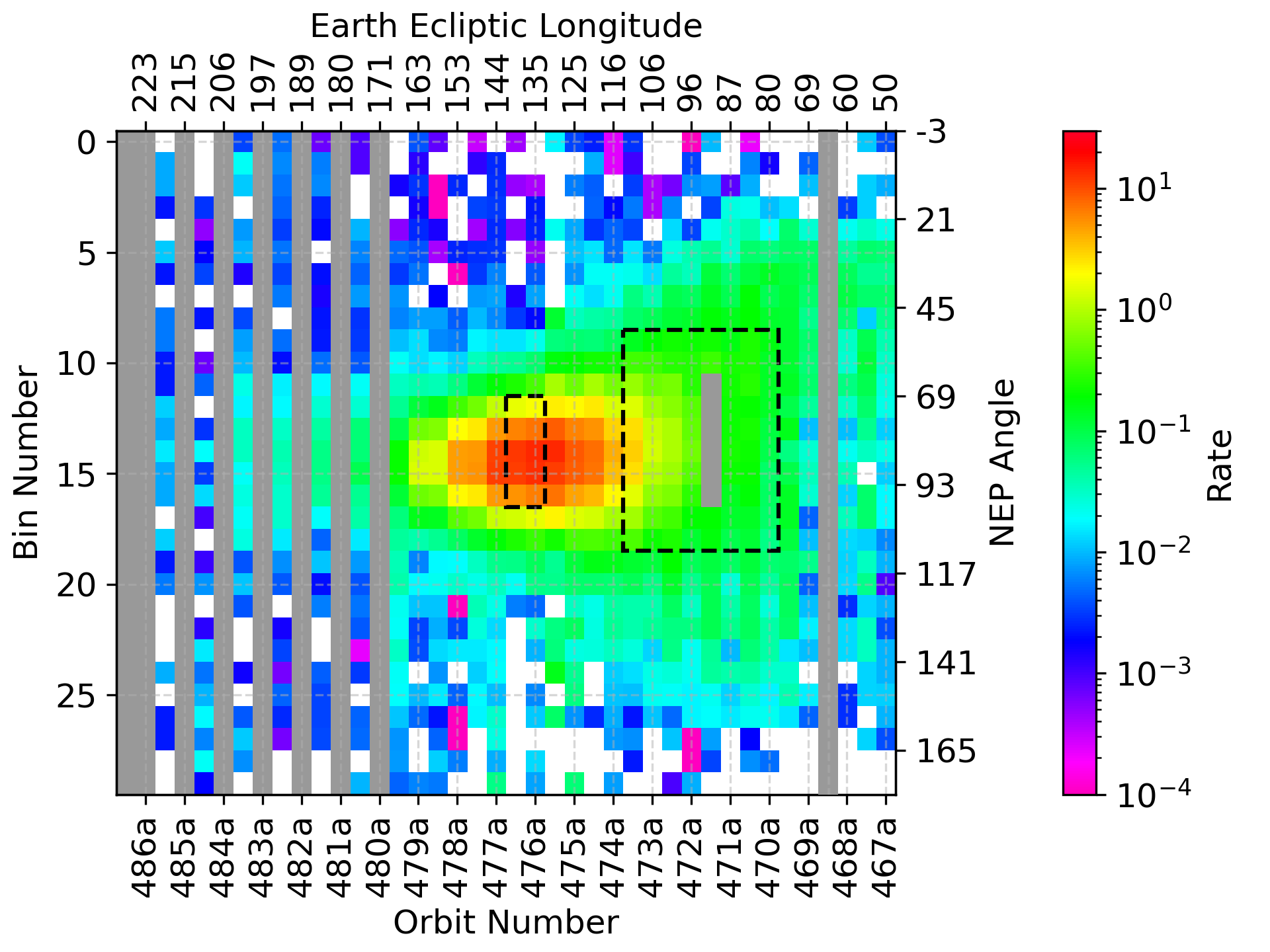}
    \caption{The data used for calculating the Renormalization Factor is presented using the ISN season 2019-2020 as an example. This plot demonstrates the selection of orbits based on Earth's Ecliptic Longitude and the NEP angle. The secondary orbits are chosen from 75$\degree$ to 115$\degree$ and bins 11 to 18, corresponding to a bin center angle range of 66$\degree$ to 108$\degree$ from the NEP. For the peak primary helium, the Earth's Ecliptic Longitude ranges from 130$\degree$ to 140$\degree$ and bins 12 to 16, corresponding to a bin center angle from 72$\degree$ to 96$\degree$ from the NEP. We have restricted our analysis to the selected region with the least amount of contribution by secondary helium. White pixels indicate a count of 0, while grey stripes denote the absence of valid good time for the selected bins (e.g., 471b) or the entire orbit (e.g., 468b).}
    \label{fig:which_data}
  \end{minipage}
\end{figure*}

 In the summer of 2012, orbit 168b, the PAC voltage was reduced to 7 kV from 16 kV  after a discharge when the spacecraft came out of an eclipse. TOF Efficiencies are lower after orbit 168, but this reduced efficiency is known based on calibration, and accounted for.  

The observations we are using in this study are in the spacecraft frame of reference and they are not Compton-Getting corrected or Survival Probability corrected, as used in  \cite{2023ApJ...954L..24G}. However there are two corrections imposed upon the observations, as briefly described here.

\textit{Throughput Correction} --- 
Before orbit 168, higher PAC voltages caused substantial unwanted TOF3 events due to background electrons. The IBEX Central Electronic Unit (CEU) has a buffer system that can store two events while another event is being processed. If another event occurs during this time, it is not registered. \citet{2015ApJS..220...26S} devised an analytical model that calculates the probability of such scenarios and provides a correction factor. This factor is a function of TOF rates monitored over six sectors (each sector covering 60 degrees) and the count rate for each 6-degree bin. Although these orbits are excluded from the calculation of the renormalization constant which will be described later, they are used to compare the width of the peak ISN helium distribution in the spin angle with later years. Further details can be found in section \ref{sec:temporal}.\\

\textit{Spin Angle offset} --- In 2016, starting with orbit 326a, a shift was observed in the ISN data after a star tracker anomaly, which essentially changed the pointing direction of the IBEX-Lo boresight by $+ 0.6^\circ$ \citet{2022ApJS..259...42S}. Before the anomaly the bin center for the $0^{\text{th}}$ bin was \(0\degree\), which changed to \(0.6\degree\) after the shift. This anomaly changes the count rate slightly in each bin from the nominal binning. To compensate for the change a redistribution method has been applied over the spin bins by fitting a Gaussian curve. More details can be found in Rahmanifard et al. [2024]

\subsection{Orbit Selection}
The science observation period for IBEX commences in early October. The secondary population of helium becomes dominant when the ecliptic longitude of the IBEX spin axis (Earth Ecliptic Longitude $+180\degree$) is approximately \(235\degree - 295\degree\) \citep{2016ApJS..223...25K}, starting in mid-November and concluding in mid-January. Consequently, for each ISN season, the secondary orbits span consecutive years; for instance, during the 2015 ISN season, the secondary population includes orbits from 2014 and 2015. Subsequently, primary interstellar helium dominates at approximately \(335\degree\) until the end of February. However, in this study, we have limited our selection of primary helium to the range of \(310\degree\) to \(320\degree\), where the maximum ISN He flux is detected. This restriction is applied to minimize the influence of secondary helium.  Observations beyond this period are focused on ISN hydrogen \citep{2019ApJ...871...52G}, extending through mid-May and covering an ecliptic longitude of approximately \(60\degree\). Orbits used in this study are given in Table \ref{tab:orbit}.

%% file: sections/3temporal_variation.tex
\section{Simulation}\label{sec:model}
The study of the ISN distribution function  began in the 1960s and early 1970s, under the assumption that the interstellar medium is cold \cite{1968Ap&SS...2..474F,1970A&A.....4..280B,1972NASSP.308..609A}. By the late 1970s, researchers such as \citet{1979A&A....77..101F}, \citet{1978AREPS...6..173T}, and \citet{1979ApJ...231..594W} started to develop hot models that account for the distribution of interstellar hydrogen at finite temperatures. Building on the hot model, \citet{2012ApJS..198...10L,2015ApJS..220...23L} proposed an analytical model designed to calculate the distribution function of ISN He within the heliosphere, specifically adapted to IBEX observations. This model utilizes an analytic formulation to estimate the survival probability of neutrals based on the ionization rate and solar wind flux, assuming that the ionization rate is constant at a specific location and decreases with the inverse square of the distance from the Sun.
 
 \citet{2013ApJ...775...86S} expanded on this analytical model by developing a numerical integration model. This model integrates the distribution function over the detailed response function of IBEX-Lo, comprising three parts: integration over the spin sector, the collimator, and energy. Detailed descriptions can be found in \citep{2013ApJ...775...86S, 2015ApJS..220...25S}.

This model is applied during the optimal ISN data acquisition times. For each three-hour interval of ISN best-time, the model is run separately, and the average count rate is calculated. At the average ISN best-time position, the real-time spin axis is determined. Using input parameters for interstellar helium and hydrogen, the count rates are then calculated. The parameters used in the model are listed in Table \ref{tab:model_para}\\
 \begin{table}[t]
\centering
\begin{threeparttable}
\caption{\\Parameters for ISN He used in the Model }
\label{tab:model_para}
\begin{tabular*}{\columnwidth}{@{\extracolsep{\fill}}lcc@{}}
\toprule
\toprule
 & \textbf{Primary He} & \textbf{Secondary He} \\
\hline
$n_\infty$ & $0.00154 \ \text{cm}^{-3}$ & $0.000878 \ \text{cm}^{-3}$ \\
$T_\infty$ & $7500 \ \text{K}$ & $9500 \ \text{K}$ \\
$V_\infty$ & $25.4 \ \text{km} \ \text{s}^{-1}$ & $11.3 \ \text{km} \ \text{s}^{-1}$ \\
$\lambda_\infty$ & $75.75^{\circ}$ & $71.57^{\circ}$ \\
$\beta_\infty$ & $5.1^{\circ}$ & $11.95^{\circ}$ \\
\hline
\end{tabular*}

\end{threeparttable}
\end{table}
\begin{figure*}[t]
  \centering
  \begin{minipage}{0.48\textwidth}
    \centering
    \includegraphics[width=\textwidth]{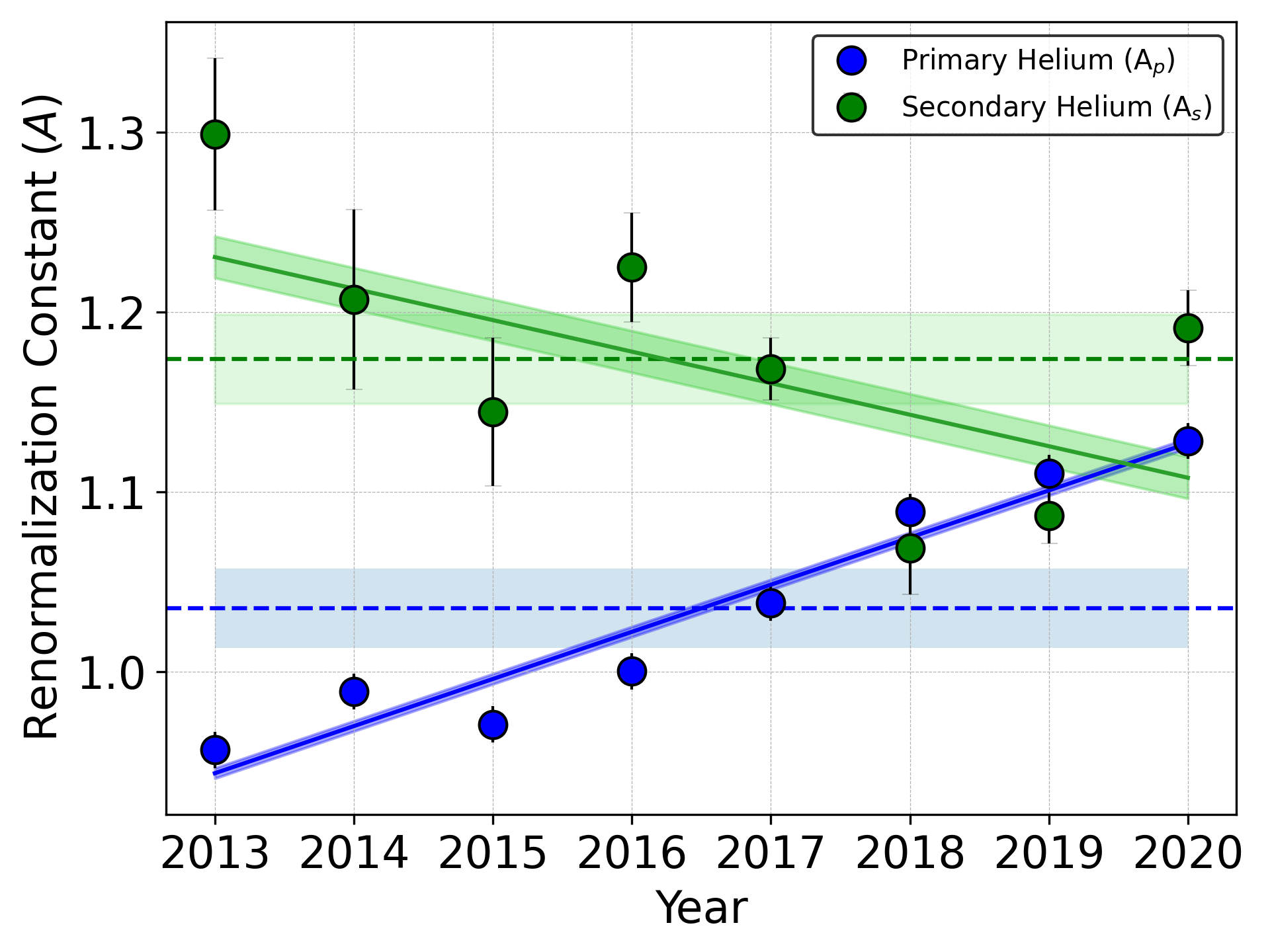}
  \end{minipage}
  \hfill
  \hfill
  \begin{minipage}{0.48\textwidth}
    \centering
    \includegraphics[width=\textwidth]{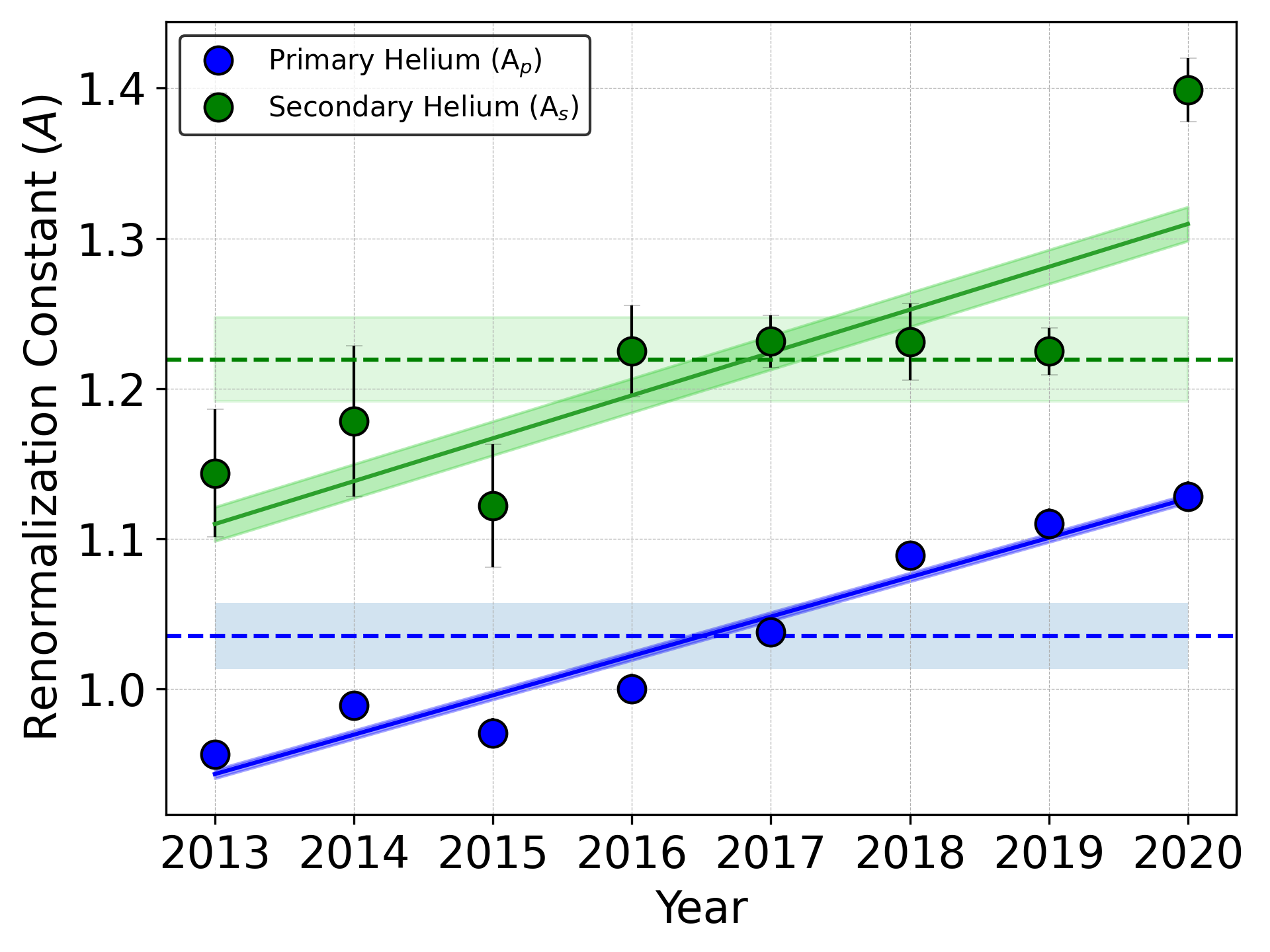}
  \end{minipage}
  
  \vspace{1em} 

  \begin{minipage}{0.48\textwidth}
    \centering
    \includegraphics[width=\textwidth]{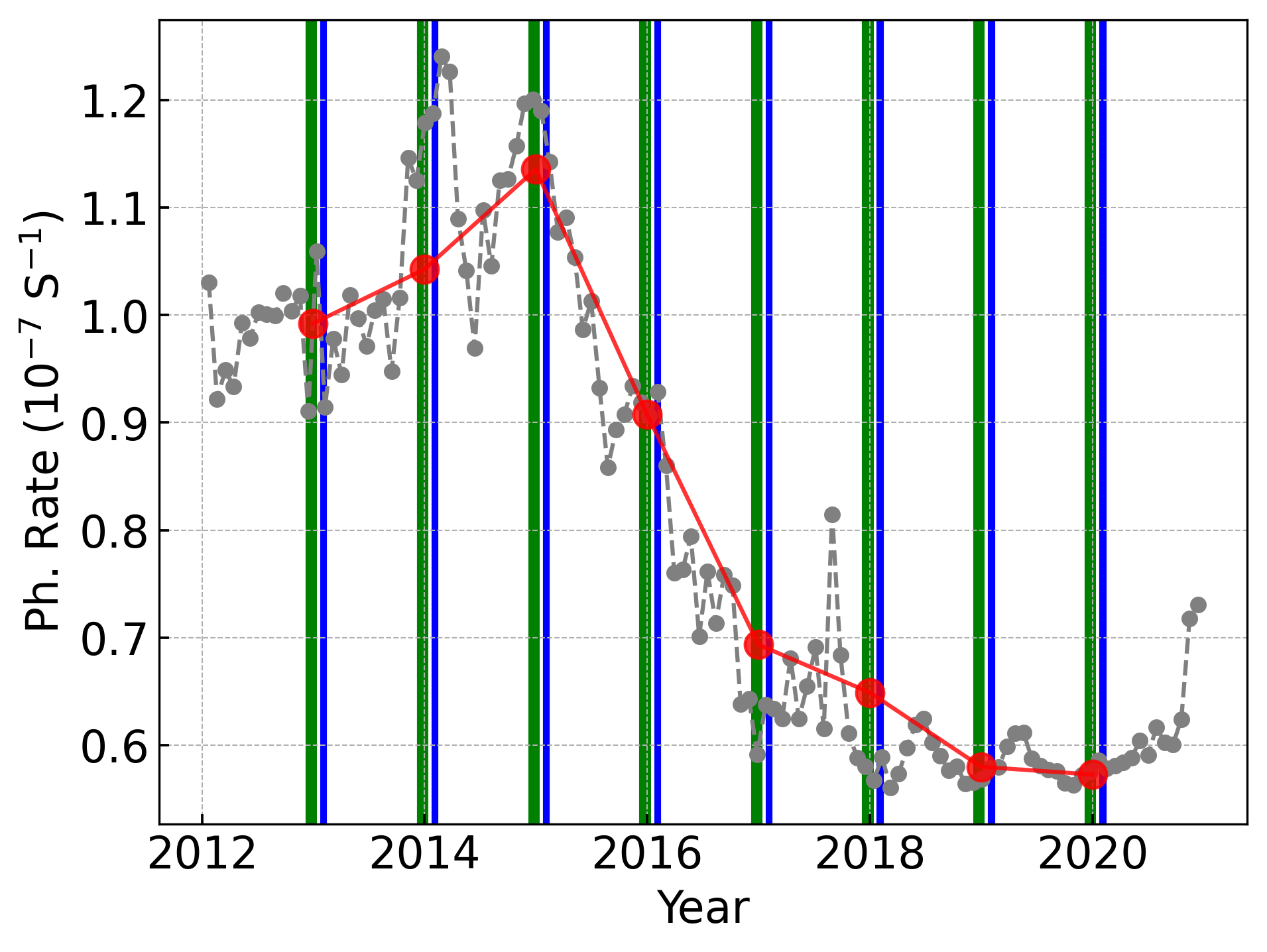}
  \end{minipage}
  \hfill
  \begin{minipage}{0.48\textwidth}
    \centering
    \includegraphics[width=\textwidth]{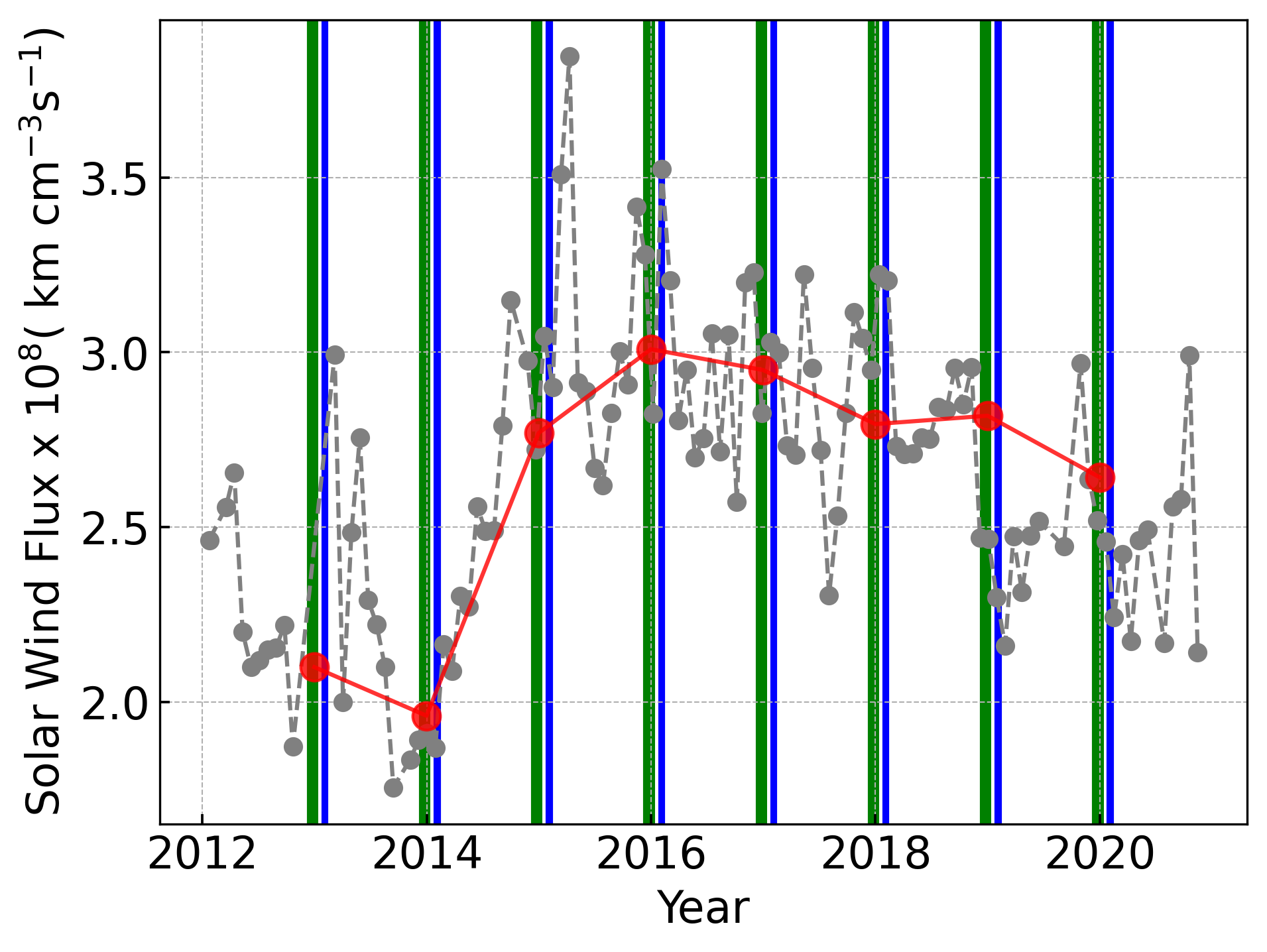}
  \end{minipage}
  
  \caption{Shown are the renormalization constants for primary ($\textbf{A}_p$) and secondary ($\textbf{A}_s$) He obtained from aFINM with the temporal variation of the photoionization rate and solar wind flux for the years 2013 to 2020. \textit{Top} -- Blue and green filled circles represent $\textbf{A}_p$ and $\textbf{A}_s$ respectively. The solid blue line represents a time dependent linear fit for $\textbf{A}_p$ which has a positive slope indicating that the renormalization constant increases with time. The dotted blue line with the light blue shaded region representing a 1$\sigma$ error, is the time independent average of $\textbf{A}_p$. Similarly the solid and dotted green lines, with their respective 1$\sigma$ error, represent time dependent and independent fits over the years. We note an important observation - the trends for the two populations are opposite in nature. \textit{Bottom left} -- This panel shows temporal variation of the photoionization rate from  2012 to 2020. Light grey dots are the photoionization rate averaged over 1 Carrington rotation of $\sim$ 27 days. The green and blue shaded regions are times for which the secondary and primary orbits have been selected in this study. The red dots show the photoionization rate averaged over 7 Carrington rotations. \textit{Bottom right }-- This panel has the same structure as the left one representing the temporal variation of the solar wind flux. We note that the photoionization rate sharply decreases after the solar maximum in 2015 but the solar wind flux is almost constant from 2015.}
  \label{fig:all_images}
\end{figure*}
\begin{figure*}[t!]
  \centering
  \begin{minipage}{0.75\textwidth}
    \centering
    \includegraphics[width=\textwidth]{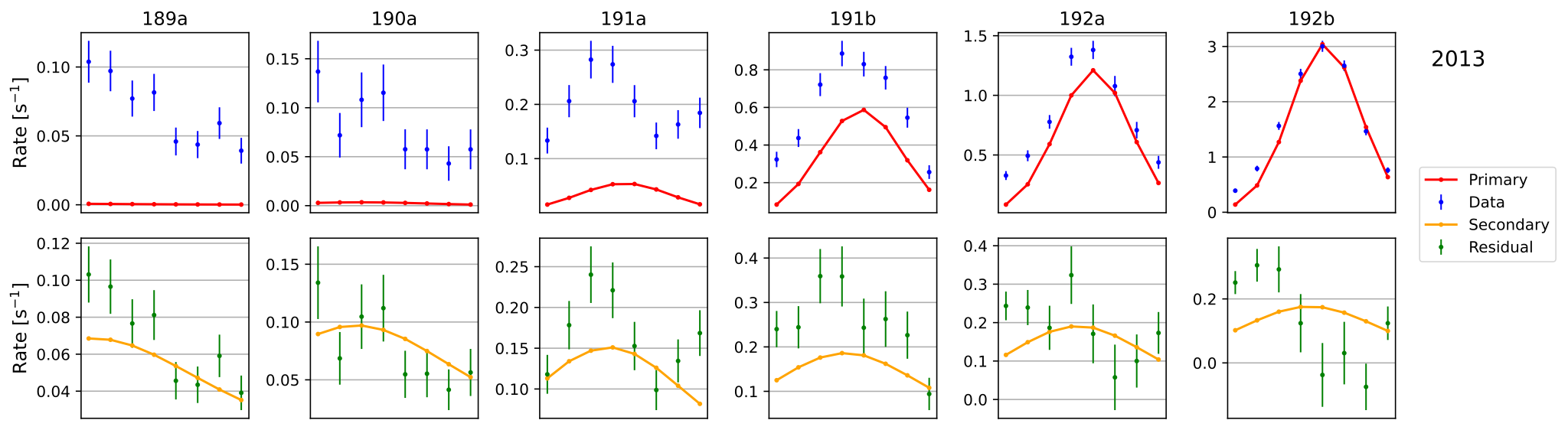}
  \end{minipage}
\end{figure*}
\begin{figure*}[t]
  \centering
  \begin{minipage}{0.75\textwidth}
    \centering
    \includegraphics[width=\textwidth]{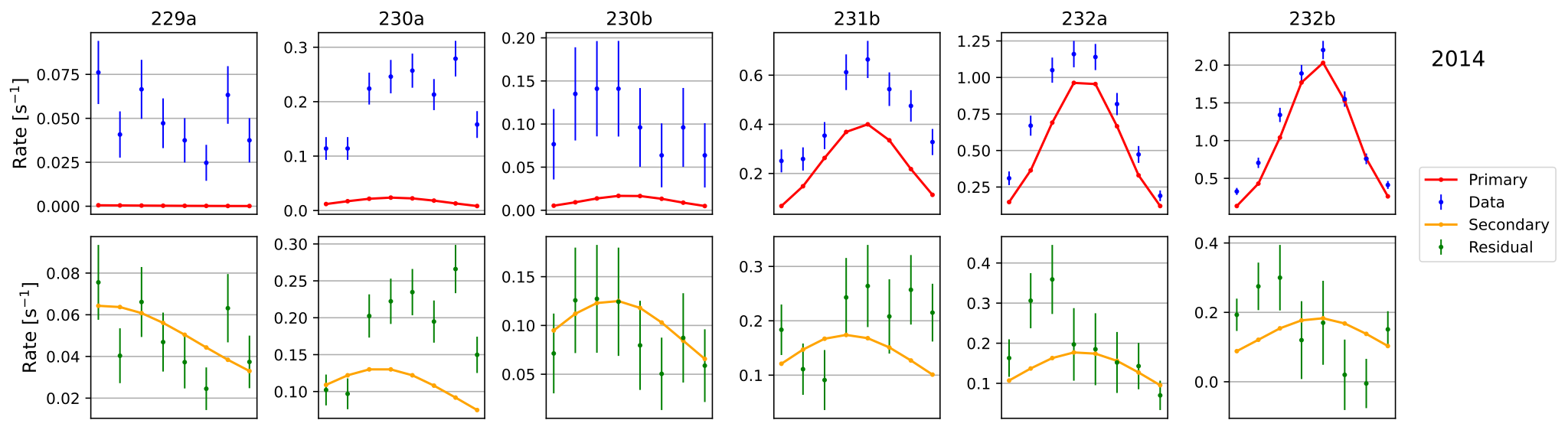}
  \end{minipage}
\end{figure*}
\begin{figure*}[t]
  \centering
  \begin{minipage}{0.625\textwidth}
    \centering
    \includegraphics[width=\textwidth]{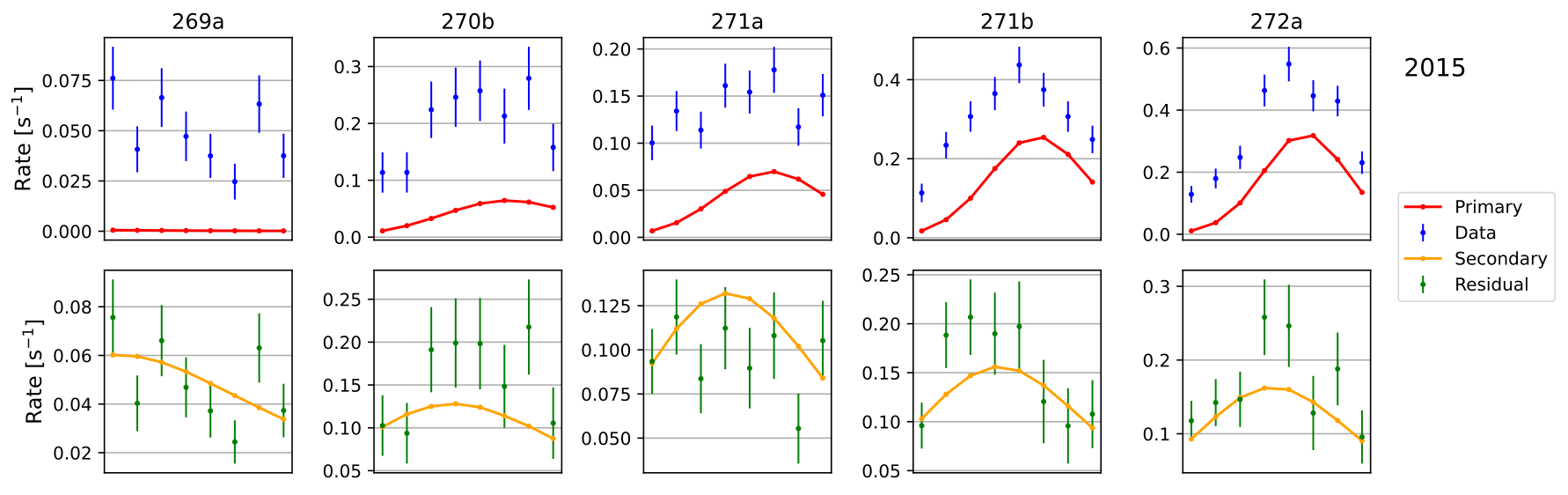}
  \end{minipage}
\end{figure*}
\begin{figure*}[t]
  \centering
  \begin{minipage}{1.0\textwidth}
    \centering
    \includegraphics[width=\textwidth]{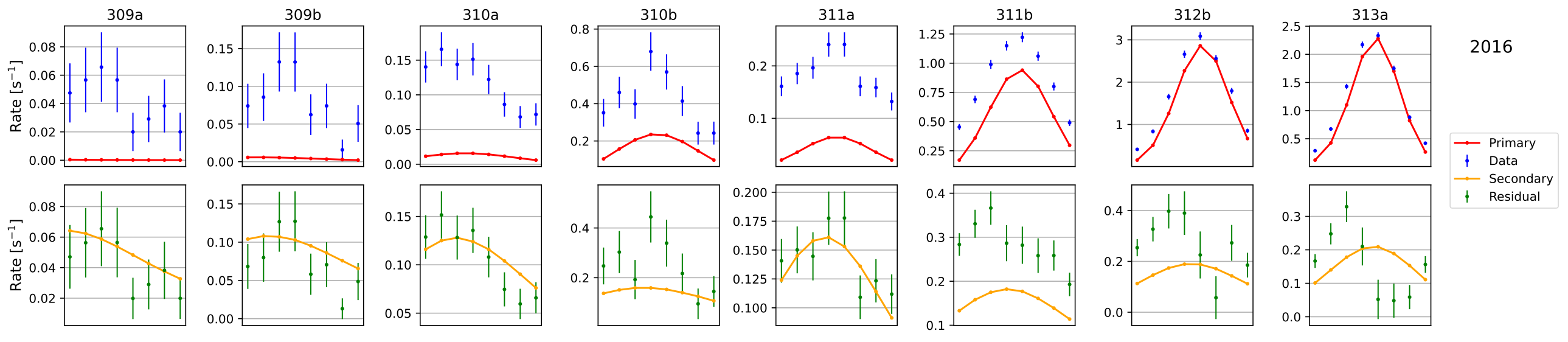}
  \end{minipage}
\end{figure*}
\begin{figure*}[t]
  \centering
  \begin{minipage}{0.875\textwidth}
    \centering
    \includegraphics[width=\textwidth]{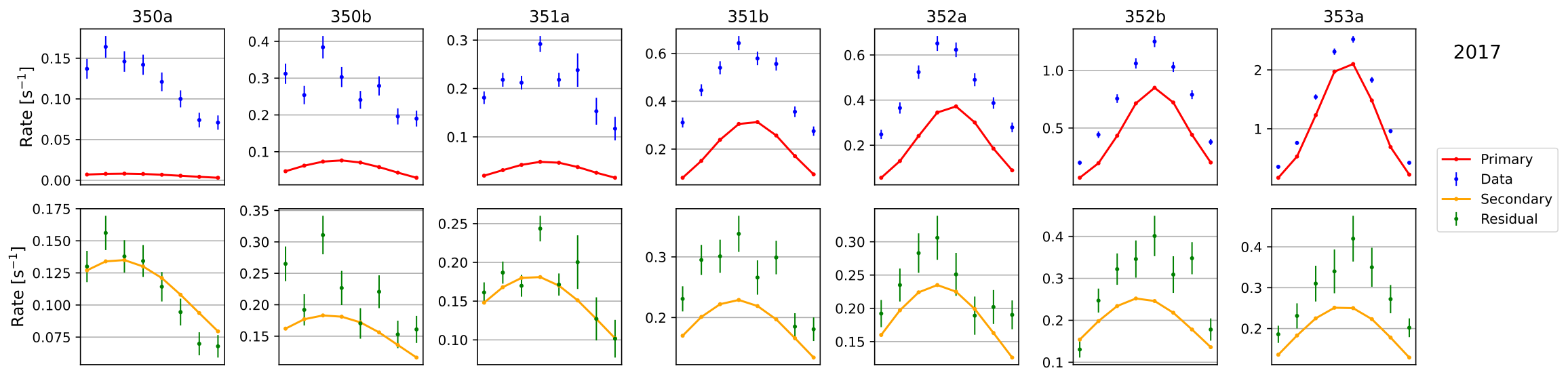}
  \end{minipage}
\end{figure*}
\begin{figure*}[t]
  \centering
  \begin{minipage}{0.75\textwidth}
    \centering
    \includegraphics[width=\textwidth]{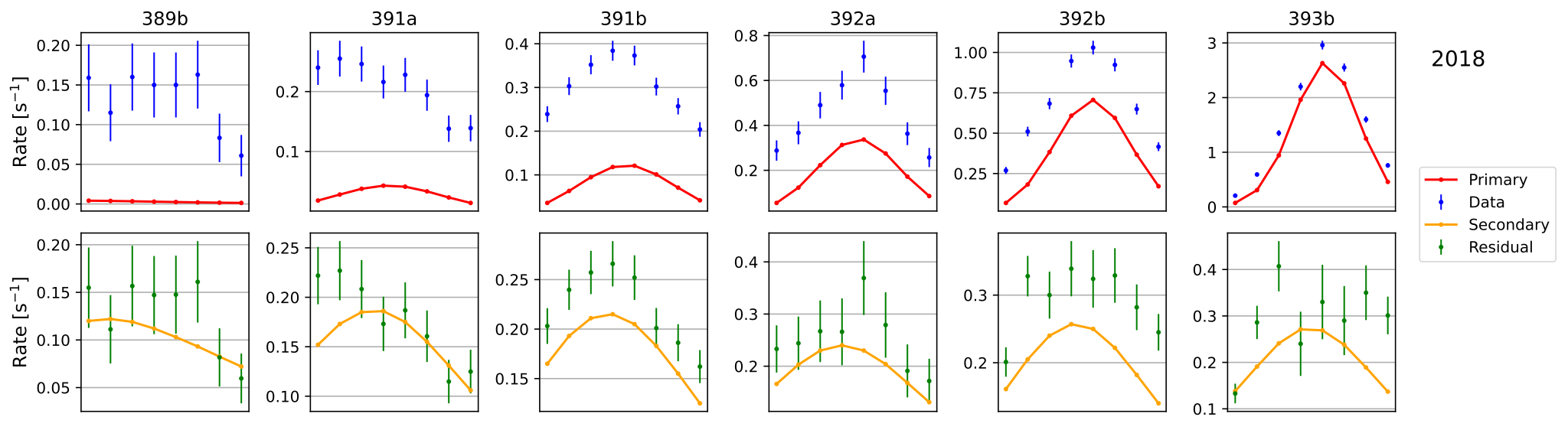}
  \end{minipage}
\end{figure*}
\begin{figure*}[t]
  \centering
  \begin{minipage}{1.0\textwidth}
    \centering
    \includegraphics[width=\textwidth]{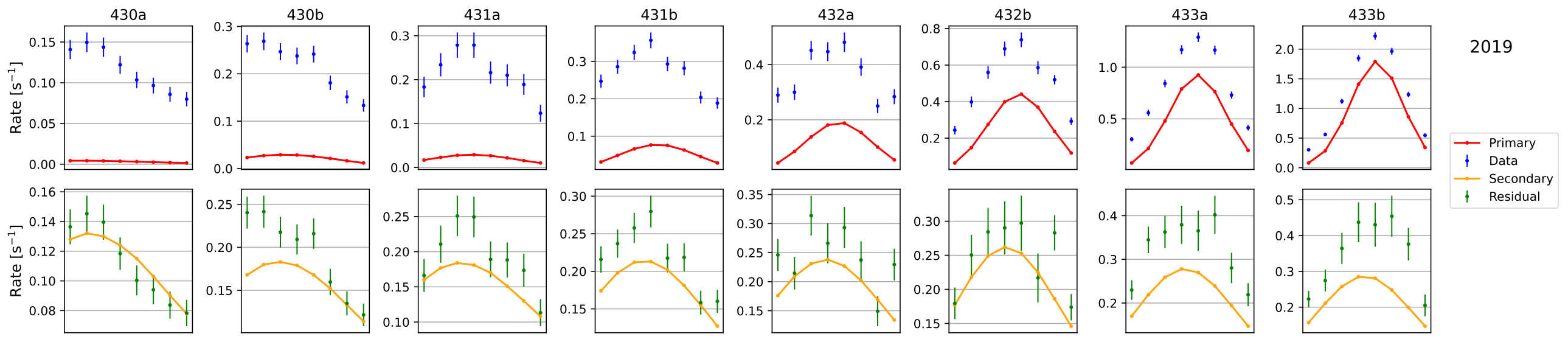}
  \end{minipage}
\end{figure*}
\begin{figure*}[t]
  \centering
  \begin{minipage}{0.875\textwidth}
    \centering
    \includegraphics[width=\textwidth]{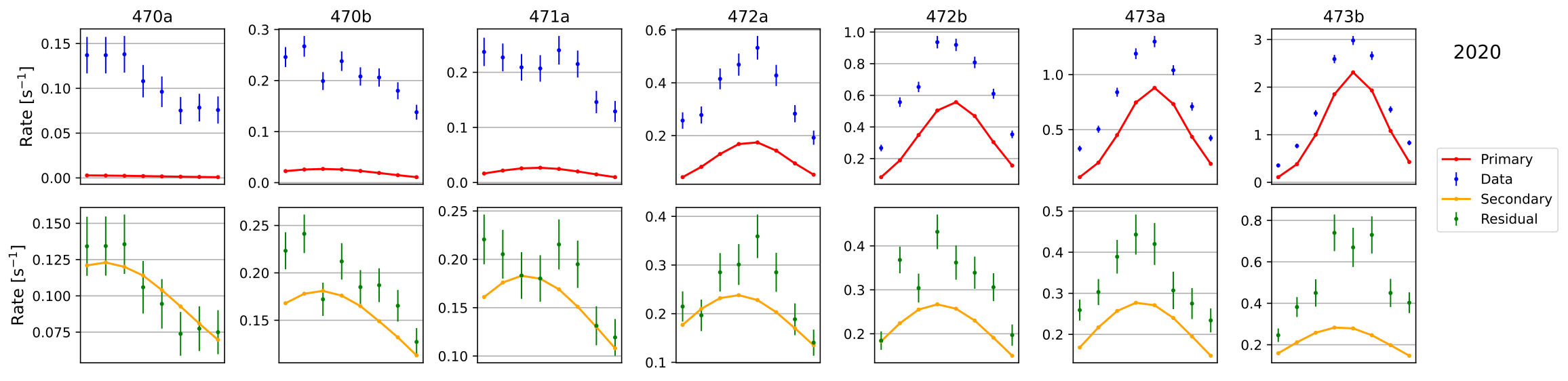}
  \end{minipage}

  \caption{Count rates in different orbits for each year that are used to estimate \( \textbf{A}_\text{s} \). Each box represents a different orbit. The blue dots in the upper panel show the count rate observed by IBEX-Lo, while the red line represents the simulated primary helium contribution. The green dots in the bottom panel show the residual after subtracting the primary helium contribution from the observations. The yellow line indicates the simulated secondary helium, which is comparable to the residual. The data points correspond to the NEP spin angle (6° bin center) from 60° to 102°.}
  \label{fig:data}
\end{figure*}

\section{Temporal Variation of renormalization constants}\label{sec:temporal}
Simulating observations by IBEX-Lo gives us an opportunity to study the modulation  of the ISNs in the vicinity of the Sun. The simulated rate  is proportional to the density of helium population ($n_{\text{He},\infty}$)  in their source region and the survival probability ($\text{S}_p$) of the atoms. The survival probability is a function of the total loss rate which varies with the activity of the Sun based on the ionization rate. An important outcome of the comparison between observation and simulation is the renormalization constant ($\textbf{A}$),
which is a factor multiplied to the modeled rate, derived in a chi-square minimum. The constant $\textbf{A}$ acts as a scaling factor which essentially modifies the intensity of the simulated rate and is proportional to $n_{\infty} \times S_p $. \\
The temporal variation of $\textbf{A}$ over the years provides direct insight into the ionization rate. Ideally, when the instrument's geometric factor, ionization rate, and density are accurately modeled, the renormalization constant should be 1. Assuming the geometric factor and density are precisely known, $\textbf{A} > 1$ indicates an underestimated ionization rate, and $\textbf{A} < 1$ suggests it is overestimated. \\
The loss of ISN He primarily due to photoionization. Extreme UltraViolet (EUV) radiation ( $\sim$ 50 nm ) from the sun is the primary source of ionization for ISN He. The intensity of the radiation varies between the maximum and minimum of each SC, over the period of approximately 11 years. We have applied an averaging scheme  over the last part of the trajectory where the loss rate is higher than 10\% of its value at 1 AU. The radius of the sphere of influence is approximately 4 AU from the sun and the time spent by ISN He inside this sphere is approximately 7 Carrington rotations of the sun \citep{2019ApJ...887..217R} or $\sim$ 190 days. We neglect ionization of ISN He by charge exchange with SW protons, as the cross section for charge exchange is very low \citep{2020ApJ...897..179S}. We also neglect electron impact ionization because the radial dependence of electron impact ionization of ISN He is not well understood \citep{1989A&A...224..290R} and is not effective beyond 2 AU \citep{2023ApJ...953..107S}.

The orbits used to estimate renormalization factors for secondary helium are shown in figure \ref{fig:data}. For each year there are two rows, while each column represents unique orbit. In the upper panel, the blue dots represent the count rate observed by IBEX-Lo, while the red line denotes the simulated contribution from primary helium. The initial orbits, which are distant from the primary helium peak, show an almost negligible contribution from the primary population. Moving from left to right, as the orbits progress, the Earth's ecliptic longitude increases, leading to a corresponding rise in the contribution from primary helium. In the lower panel, the green dots show the residuals after subtracting the primary helium contribution from the observations. The yellow line represents the simulated secondary helium, which should aligns with the residuals.
We calculate $\textbf{A}_\text{P}$ for primary helium and $\textbf{A}_\text{S}$ for secondary helium each year from 2013 to 2020. The reason for choosing these years are the following: (a) In summer of 2012 the Post Acceleration (PAC) Voltage was reduced to 7 kV from 16 kV. Due to the PAC voltage change, the efficiency of IBEX-Lo is reduced nearly by a factor of two \citep{2023ApJ...953..107S}, which introduces an inconsistency into the temporal variation of $A$. (b) In the orbits where primary He is dominant, the contribution of secondary helium is negligible (the ratio of primary to secondary is approximately 30). However, the opposite is not true, thus we need to subtract the contribution of the primary population in the orbits dominated by the secondary helium. We will discuss later the effect of $\textbf{A}_\text{P}$ on calculating $\textbf{A}_\text{S}$. In 2011 and 2012 the primary helium data statistics is not substantial with the ISN best-time we have used in this study, subsequently these years are not included into the analysis. (c) Before orbit 130 in 2011, the orbits are longer in period $\sim$  7.5 days. From orbit 130 the orbits are divided into arcs, each with a duration of  $\sim$ 4.5 days. However, this change does not affect the analysis. 

The top two panel of Figure \ref{fig:all_images} shows the temporal variation of  $\textbf{A}_\text{p}$ and $\textbf{A}_\text{s}$.  $\textbf{A}_\text{p}$  remains almost constant from 2013 to 2015  (solar maxima) and gradually increases onwards.  A time-dependent linear fit using years as a variable yields a positive slope of \( +0.02 \). The average value of \( \textbf{A}_\text{p} \) is found to be \( 1.04 \pm 0.02 \), indicating that the geometric factor used in the model is very close to the actual value. It is important to note that this geometric factor was derived assuming a primary helium density at infinity of \( n_\infty = 0.0154 \, \text{cm}^{-3} \).  \citet{2023ApJ...953..107S} report similar findings from year 2009 to 2020 and argue that an overall 40\% increase in the photoionization rate could explain this trend. However, the total uncertainty in the photoionization rate is at most 20\%, rendering this assumption invalid. Another possibility considered was that the efficiency of IBEX-Lo is increasing over time, which is unrealistic. An unknown source at solar minima which is not present at the solar maxima may also explain the trend. The authors ultimately concluded that they could not resolve the matter.\\

As mentioned earlier, in peak primary orbits, the contribution of secondary helium is negligible, but this does not hold for secondary orbits. In the top left panel, $\textbf{A}_\text{s}$ is calculated by multiplying $\textbf{A}_\text{p}$ with the simulated primary helium rate before subtracting it from the data, whereas in the right panel, $\textbf{A}_\text{s}$ is calculated without multiplying by $\textbf{A}_\text{p}$. The trend of $\textbf{A}_\text{s}$ in both panels is completely opposite, highlighting the strong influence of $\textbf{A}_\text{p}$.

Unlike the primary helium, the renormalization constant for secondary helium, $\textbf{A}_\text{s}$, in the left panel shows a negative slope of $-0.01 \pm 0.01$, indicating a systematic decrease starting in 2013. The time-independent average value of $\textbf{A}_\text{s}$ is significantly higher at $1.15 \pm 0.02$, suggesting that the count rate of secondary helium is, on average, 15\% greater than predicted by the model. \citet{2022ApJS..261...18G} also reported unexpectedly high ENA intensity in 2015. Although the connection between this ENA surge and ISN helium is not well established, it may be relevant.

In contrast, when $\textbf{A}_\text{s}$ is calculated without including $\textbf{A}_\text{p}$, the trend aligns more closely with that of primary helium, with a slope of $0.03 \pm 0.01$ and a time-independent average of $1.22 \pm 0.03$. Notably, $\textbf{A}_\text{s}$ shows a sudden jump after 2015, remaining constant until 2019, with a substantial increase in 2020 compared to previous years.

The temporal variations of the photoionization rate and solar wind proton flux in the ecliptic plane at 1 AU are shown in the bottom left and right panels of Figure \ref{fig:all_images}. The photoionization rate was significantly higher during 2013–2016 compared to subsequent years, and $\textbf{A}_\text{p}$ reflects this trend, suggesting an underestimation of ionization loss. However, during this period, $\textbf{A}_\text{s}$ in the left panel was higher than in 2018 and 2019, which seems counterintuitive. In the right panel, $\textbf{A}_\text{s}$ increased after 2015 but remained constant, which also does not follow the photoionization rate trend.

An additional contribution to the secondary population during periods of higher solar activity could explain this anomaly. We suggest that the scattering of primary ISN helium by solar wind protons may be responsible for this increase. The bottom right panel of the figure shows the solar wind flux from 2012 to 2020, with a rapid increase in mid-2014 followed by a slow decrease. A visual inspection of this trend alongside the temporal variation of $\textbf{A}_\text{s}$ reveals a correlation. In the next section, we will discuss the scattering mechanism and its correlation with solar wind flux. \\

%% file: sections/4scattering.tex
\section{Scattering Inside Heliosphere}\label{sec:scattering}
ISN helium experiences loss primarily due to photoionization by solar extreme ultraviolet (EUV) radiation, a well-understood process. A minor contribution to this loss also arises from charge exchange and electron impact ionization \citep{1989A&A...224..290R,2013A&A...557A..50B,2014ApJS..210...12B}. However, another physical process alters the ballistic trajectory of ISN He, causing reduction of flux in the core of the distribution: elastic scattering by the solar wind. \citet{1964Icar....3..253B} first highlighted the significance of interstellar helium scattering by the solar wind. \citet{1975P&SS...23..419W}, \citet{1979A&A....77..101F} and \citet{1974spre.conf..567F} realized that the temperature of interstellar helium and hydrogen ought to increase by collisional heating. Originally this heating mechanism was treated as a continuous process of energy and momentum exchange inside the heliosphere \citep{1974MNRAS.167..103W,1975P&SS...23..419W,1979ApJ...231..594W}
In 1986, \citet{1986P&SS...34..387G} pointed out that, given the number of collisions an interstellar neutral atom experiences throughout its trajectory, continuous energy transfer is not applicable in this scenario. He also concluded that most atoms will gain very little momentum, resulting in a very thin but long wing of the distribution. A quantitative assessment of this theory was not provided until 2013 \citep{https://doi.org/10.1002/jgra.50199}. Based on that work we  argue that the increase of the secondary helium flux during solar maximum can be explained by the scattering of primary ISN He by SW protons for the following reason: In the ecliptic plane, IBEX-Lo observes the peak for ISN He at a ``sweet spot"  that occurs at $130\degree$, measured from the Vernal Equinox in the downwind direction. The relevant angular range where scattered particles distinguish themselves from the peak extends from less than $100\degree$ to more than approximately $160\degree$. The region near $100\degree$ is primarily influenced by secondary helium. The area around $160\degree$ is dominated by interstellar hydrogen and we predict similar increase of ISN hydrogen flux in that region.\\

The author also predicts that similar halo would also be seen in the plane perpendicular to the ecliptic plane. The top panel of Figure 5 in \citep{https://doi.org/10.1002/jgra.50199} shows three curves, the blue curve represents the spin angle distribution without any collision, green is for multiple elastic collisions, and red for a single elastic collision assumption, which has been used in that study. The middle and bottom panels show the ratio of one-to-no-collision with the no-collision model. 

As IBEX spins, it completes a full rotation approximately every 15 seconds, sweeping its field of view across a wide swath of the sky. A new spin starts when IBEX-Hi points toward $-3\degree $ from the North Ecliptic Pole (NEP), which is $177 \degree$ for IBEX-Lo. The primary ISN He flux core is observed at approximately $264 \degree$. Theoretically, the collision-produced halo dominates at angles larger than $30\degree$ to $35\degree$ from the core of the helium flux which corresponds to angles less than $235\degree$ and more than $295 \degree$ in the IBEX-Lo swath.
To reproduce similar halo perpendicular to the ecliptic plane, we chose all the primary orbits used to calculate $\text{A}_\text{P}$and subtracted respective secondary helium  from the observations. The simulated rate for primary helium  serves as a conventional model without collision and the subtracted observation as the scattered distribution. 

\begin{figure}
    \centering
    \includegraphics[width=1\linewidth]{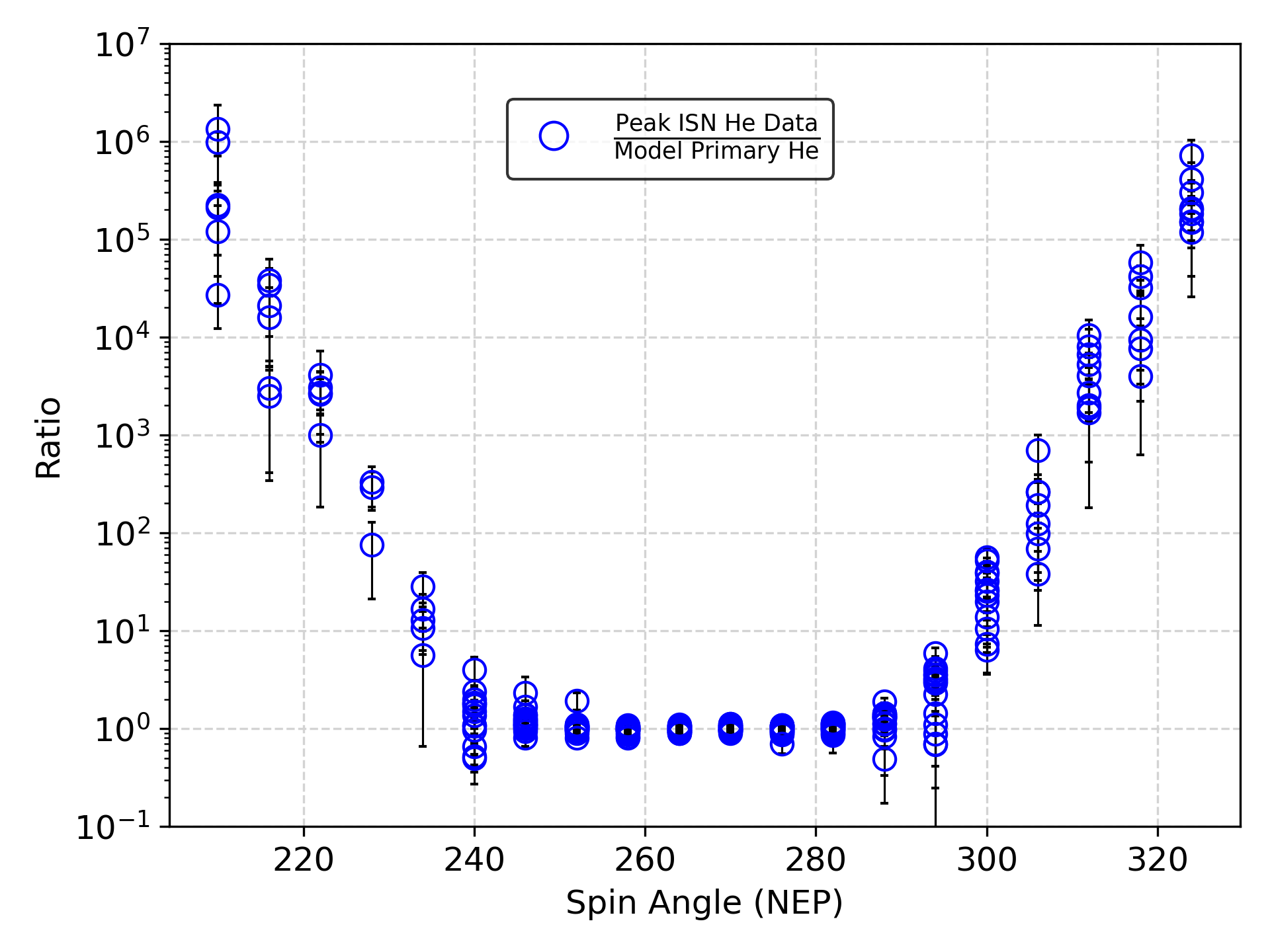}
    \caption{The helium flux halo observed by IBEX-Lo, which can be compared to the predicted ratio by  \citet{https://doi.org/10.1002/jgra.50199}. The primary helium orbits utilized in this study correspond to those specified in Table \ref{tab:orbit}. Refer to the accompanying text for further details.}
    \label{fig:halo}
\end{figure}

The theoretically predicted halo is also observed by IBEX-Lo as seen in Figure \ref{fig:halo}. The ratio between  observation with the simulation is comparable to the middle panel of the said figure. Here we notice that at spin angle around 220 degrees the ratio is about 20 but the observed ratio by IBEX-Lo is of the order of thousand, 50 times higher than that theoretically predicted ratio. This deviation can be explained by using a higher scattering cross section. In the analytical model scattering cross section between Helium and solar wind protons is assumed to be $0.56\times 10^{-16} \text{cm}^2$ at $450 \text{km/s}$ with density $5 \text{cm}^{-3}$.
In reality the solar wind density is higher and its speed fluctuates between 350-800 km/s. More specifically, the cross section falls off with increasing relative velocity between the interacting particles. Also, when the relative speed is lower the resulting scattering angle is higher. Thus, a higher cross section with higher scattering angle from a relative lower speed may explain the deviation but that analysis is beyond the scope of this work. \citep{1986P&SS...34..387G}.\\

\begin{figure}
    \centering
    \includegraphics[width=1\linewidth]{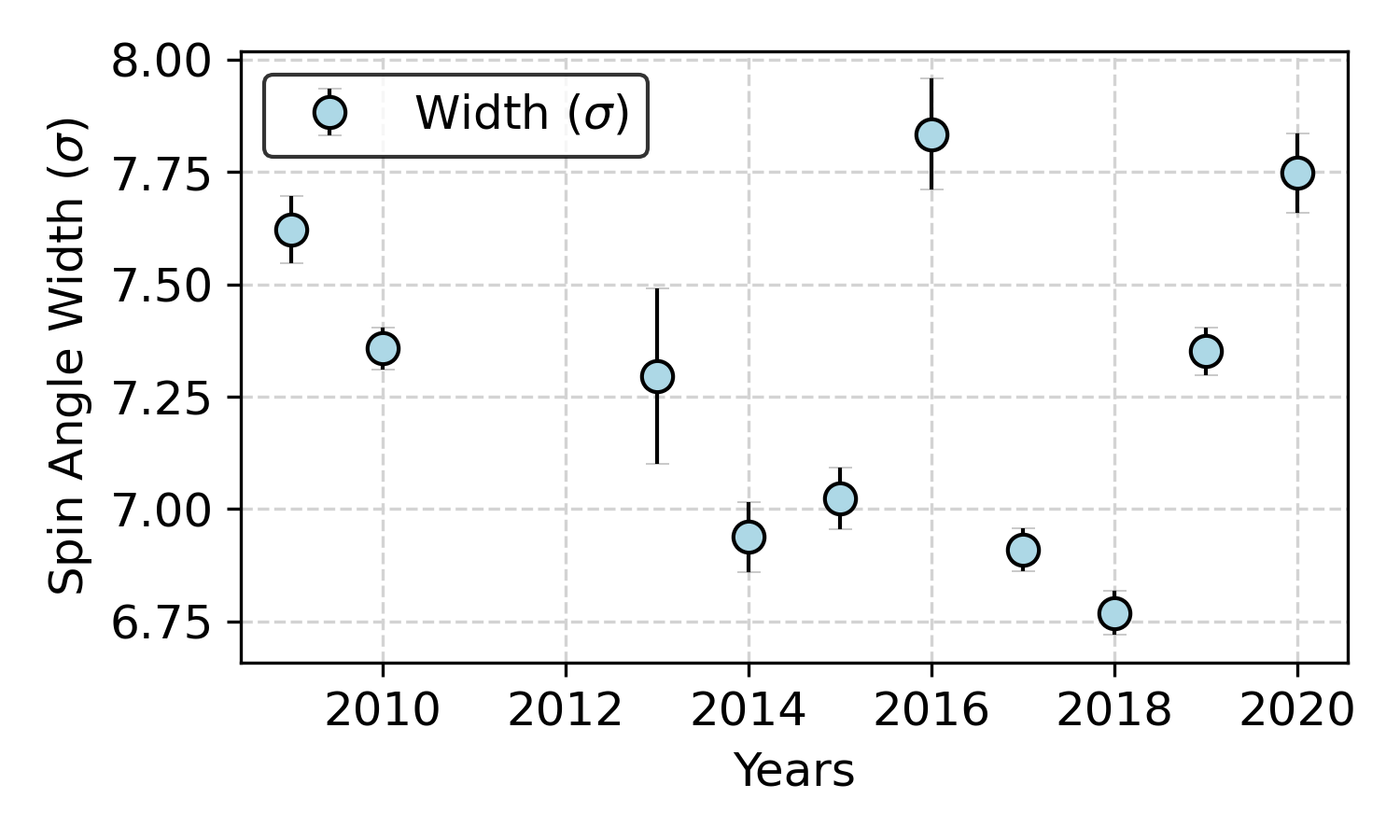}
    \caption{Width ($\sigma$) of the primary ISN He distribution in the latitudinal plane. The width is the standard deviation of the Gaussian Distribution fitted for the spin angle range 60$\degree$ to 120$\degree$. }
    
    \label{fig:sigma}
\end{figure}
 In SC 24 the SW flux increased by 50\% in 2015 over the previous years and decreases slowly. It is intuitive to anticipate that the scattering phenomena will result in a broadening of the particle distribution in the spin-angle plane perpendicular to the ecliptic plane when the SW flux is higher.  However, observations from IBEX-Lo reveal an inverse tendency, as shown in Figure \ref{fig:sigma}. With the notable exception of 2016,2017,2018 particle distributions exhibit a narrowing effect during the solar maximum (2013,2014,2015) and a broadening effect during the solar minimum (2009, 2010, 2019, and 2020).A possible explanation is that, during solar maximum, the solar wind flux at higher latitudes is greater than in the ecliptic plane, leading to increased charge exchange between helium atoms and solar wind particles \citep{2020ApJ...897..179S}. Additionally, the photoionization rate may be higher at elevated ecliptic latitudes during periods of high solar activity, further reducing the population of neutral particles available for scattering and subsequent detection. To verify these assumptions, a 3D ionization model that incorporates the ionization rate along with the trajectory of each particle is required which is beyond the scope of this work. The atypical broadening observed in 2016 can possibly be attributed to  statistical anomalies \citep{2022ApJS..259...42S}.

 \subsection{Velocity Dependent Cross Section}
The elastic scattering cross section of proton and Helium atom is a function of energy. With increasing relative speed the cross section and the scattering angle both decreases. We have used the cross section data released by \citet{2021ApJ...911L..36S}. The relevant speed range for the interaction between SW proton and helium is about 200 km/s to 1000 km/s. In this range we fit a power law equation 
\begin{equation}
\sigma_{sc}=3978.551 \times V_r^{-1.432}
\label{eq:sigma}
\end{equation} to interpolate the scattering cross section for any relative speed, $V_r$. Based on the previous work by \citet{https://doi.org/10.1002/jgra.50199}, we have assumed only 10\% of the interaction leads to a scattering angle more than 1$\deg$ which is responsible for the halo formation. 
\subsection{Correlation with the solar wind flux}
The observation of halo formation in primary helium orbits and the increased flux in secondary helium orbits confirm that the scattering effect is real and modifies IBEX observations. The temporal variation of secondary helium should exhibit a correlation between the scattering rate and the renormalization factors.

This correlation does not apply to the primary renormalization factor because the count rate for primary helium is significantly higher than the loss due to scattering, and uncertainties in the photoionization rate play a dominant role. When calculating $\textbf{A}_\text{S}$ subtracting the primary contribution from the data after multiplying it by $\textbf{A}_\text{P}$, assuming the same modulation applies across different orbits. However, the correlation between the scattering rate and $\textbf{A}_\text{S}$ was found to be -0.15, indicating almost no correlation. In contrast, when we subtract the primary contribution without multiplying by $\textbf{A}_\text{P}$, the correlation with the 452 days average scattering rate prior to the observation period is  $\sim$ 0.76, showing a strong correlation. This similarity between the renormalization factors and the scattering rate is visible in the Figure \ref{fig:all_images}.

%% file: sections/5density.tex
\section{Revised Estimation of Interstellar Helium Density} \label{sec:density}
Estimating the density, temperature, and velocity vector of ISN helium has been a key focus in interstellar medium studies over the past few decades. While the temperature and velocity vector of the ISN helium flow can be determined from the observed angular distribution, accurate instrument calibration is essential to determine the density in the pristine interstellar medium. Additionally, understanding the various filtration processes affecting helium atoms as they travel from the pristine interstellar medium to 1 au is fundamental. In a typical forward modeling approach, an initial density in the pristine interstellar medium is assumed, followed by applying the instrument response and accounting for the loss rate of the incoming flow. \\
\comment{
\begin{table}
    \centering
    \caption{\\Loss Rate of ISN He by Different Mechanism in SC 24}
    \begin{tabular*}{\columnwidth}{@{\extracolsep{\fill}}lcc@{}}
        \toprule 
        \toprule
         Loss Mechanism&  Maximum& Minimum\\
         \hline
         Photoionization&  $12 \times 10^{-8}$& $6 \times 10^{-8}$\\
         Scattering&  $1.7 \times 10^{-8}$& $1.2 \times 10^{-8}$\\
         Electron Impact&  $1.5 \times 10^{-8}$& $1.2 \times 10^{-8}$\\
         Charge Exchange&  $0.37 \times 10^{-8}$& $0.1 \times 10^{-8}$\\
         \bottomrule
    \end{tabular*}
    
    \label{tab:ionization}
\end{table}
}%
The loss rate of helium atoms from the flux core due to elastic scattering by SW proton at heliocentric distance $r$ is,
\begin{equation}
\beta_{sc}(r)=n_{\text{p}}(r) v_{\text{p}}(r) \sigma_{sc}   
\end{equation}
where $n_{\text{p}(r)} \text{ and } v_{\text{p}(r)} $ are the density and velocity of the solar wind protons and $\sigma_{sc}$ is the scattering cross section. Loss rate by photoionization and elastic scattering both vary with the inverse square of the heliocentric distance.  The total loss rate due to photoionization $\beta_{ph}$ and scattering $\beta_{sc}$ at any $r$ is,
\begin{equation}
 \beta(r) = \big(\beta_{ph}(r_E) + \beta_{sc}(r_E)\big ) \left ( \frac{r_E}{r}^2\right ) 
\end{equation}
where $r_E$ is the distance of the Earth from the Sun, \textit{i.e.} 1 au.

The intensity of the helium count rate \text{R} obtained by forward modeling  is directly proportional to the geometric factor of the instrument \text{G}, the density of helium in the pristine interstellar medium ($\text{n}_{He,\infty} $) and the survival probability of the neutral helium along its trajectory ($\text{S}_p$),
\begin{equation}
    R \propto G \times \text{S}_p \times \text{n}_{\text{He},\infty}.
\end{equation}
The survival probability of a neutral atom \citep{2003AnGeo..21.1315R} is defined by 
\begin{equation}
    S_p(r(t)) = \exp{\left[-\int_{t_s}^{t_r} \beta(r(t),t) \, dt\right]},
\end{equation}
where $t_s$ is the starting time when ionization is relevant, $t_r$ is the time when the atom reaches $r$, certainly $t_s < t_r$. $\beta(r(t))$ is total ionization rate at heliocentric distance $r$. The analytical solution \citep{2012ApJS..198...10L} of the above equation assuming a constant ionization rate at distance $r$ from the sun  is 
\begin{equation}
    S_p(r) = \exp{\left[-\frac{\beta(r)r^2\Delta\theta}{L}\right ]},
\end{equation}
where $L$ is the angular momentum and $\Delta\theta$ is the angle swept out by the atom from infinity to $r$. A change in loss rate affects the estimation of $\text{S}_p$ and subsequently $n_{\text{He},\infty}$. In the simulation of different loss rates, the same count rate $R$ can be obtained by simultaneously adjusting ${S}_p \text{ and } n_\infty$. From Equation 2 it is straightforward to obtain,
\begin{equation}
    n_{\text{He},\infty} ^{2}=n_{\text{He},\infty} ^{1} \frac{S_p^1}{S_p^2},
\end{equation}
which reduces to 
\begin{equation}
    n_{\text{He},\infty} ^{2}=n_{\text{He},\infty} ^{1} \exp{\left[\frac{r^2\Delta\theta \Delta \beta}{L}\right ]},
    \label{eq:mfactor}
\end{equation}
where $\Delta \beta$ is the change in the loss rate. An increase in the loss rate suggests $\Delta \beta > 0$ resulting $n_{\text{He},\infty} ^{2} > n_{\text{He},\infty} ^{1}$ and vice-versa. 

\begin{table*}[t]
\centering
\caption{\\Revised Density of  Interstellar Neutral Helium in the Pristine Interstellar Medium [$\text{n}_{\text{He},\infty}$] for the inclusion of Scattering Loss and Warm Breeze Production}
\label{tab:result}
\begin{tabularx}{\textwidth}{ccccc}
\toprule
\toprule
\begin{tabular}{@{}c@{}}\textbf{Previous} \\ \textbf{Studies}\end{tabular} & 
\begin{tabular}{@{}c@{}}\textbf{Density} \\ \textbf{($\text{cm}^{-3}$)} \\ {[$\textbf{n}_{\text{He},\infty}$]}\end{tabular} &
\begin{tabular}{@{}c@{}} \textbf{Correction} \\ Scattering \\ Warm Breeze \end{tabular} & 
\begin{tabular}{@{}c@{}}\textbf{Revised Density } \\ \textbf{($\text{cm}^{-3}$)} \\ {[$\textbf{n}_{\text{He},\infty}$]}\end{tabular} & 
\begin{tabular}{@{}c@{}}\textbf{Methodology} \\ \textbf{(Instrument)}\end{tabular} \\

\midrule
\text{\citet{2004A&A...426..855V}} & $0.013 \pm 0.003$ & NA & $0.013 \pm 0.003$ (NA) & \begin{tabular}{@{}c@{}}UV Backscatter \\ (EUVE)\end{tabular}  \\
\hline
\text{\citet{2004A&A...426..845G}}& $0.0151 \pm 0.0015$ & NA & $0.0151 \pm 0.0015$ (NA) & \begin{tabular}{@{}c@{}}Pickup Ions \\ \text{(Ulysses SWICS)}\end{tabular} \\
\hline
\text{\citet{2004A&A...426..835W}} & $0.015 \pm 0.0028$ & \begin{tabular}{@{}c@{}} $1.034 \pm 0.002$ \\ $1.057 \pm 0.004$\end{tabular} & $ 0.0164 \pm 0.0032 $ & \begin{tabular}{@{}c@{}} Direct Detection \\ (Ulysses GAS)\end{tabular} \\
\hline
\text{\citet{2015ApJ...801...62W}}${ }^1$ & $0.0162 ^{+0.0058}_{-0.0031}$ &\begin{tabular}{@{}c@{}} $1.018 \pm 0.001$ \\ $1.057 \pm 0.004$\end{tabular} & $0.0174 ^{+0.0062}_{-0.0033}$ & \begin{tabular}{@{}c@{}} Direct Detection \\ (Ulysses GAS)\end{tabular} \\
\hline
\text{Combined Revised Density ${}^2$} &  $0.0148 \pm 0.0020 $ &NA & $0.0153 \pm 0.0011 $ & $\cdot $ \\
\bottomrule
\end{tabularx}
\begin{minipage}{\textwidth}
\footnotesize
${ }^1$ Estimated only for 2006-2007.\\
${ }^2$ Calculated using only top three studies.
\end{minipage}
\end{table*}
Inside the heliosphere, the scattering of ISN helium by SW protons is a source for the redistribution of ISN helium from the  core of the distribution, which reduces the intensity of the observed flux. Thus, this mechanism should be accounted for along with the other ionization loss of He, effectively increasing the total loss rate. The exponential term in equation \ref{eq:mfactor} represents the correction factor that adjusts the previously estimated density. In this work, we assume the bulk velocity of helium atoms in the pristine interstellar medium to be approximately 26 km/s, allowing us to calculate the angular momentum \( L \) and the angular deflection \( \Delta\theta(r) \). With these values, the correction factor can be straightforwardly determined.  We emphasize that scattering does not result in an actual loss of helium atoms, rather the atoms are redistributed far enough from the flux core.  

The three primary methods used for studying the interstellar medium are: (1) EUV resonant backscattering at 584 Å for neutral helium (2) observation of $\text{He}^+$ pickup ions, and (3) direct detection of ISN helium by the GAS experiment onboard Ulysses. A coordinated effort by \citet{2004A&A...426..897M} to combine all three methods resulted in a density estimate of \( n_{\text{He},\infty} = 0.0148 \pm 0.0020 \). The density estimated by the above mentioned studies assumed no filtration of neutral helium flow at the outer heliosheath creating warm breeze which was first discovered in 2014. Below, we will analyze which of the three  methods requires revision to account for scattering and the Warm Breeze.

\subsection{Resonant Backscattering} 
\begin{figure}[t]
  \centering
    \centering
    \includegraphics[width=1\linewidth]{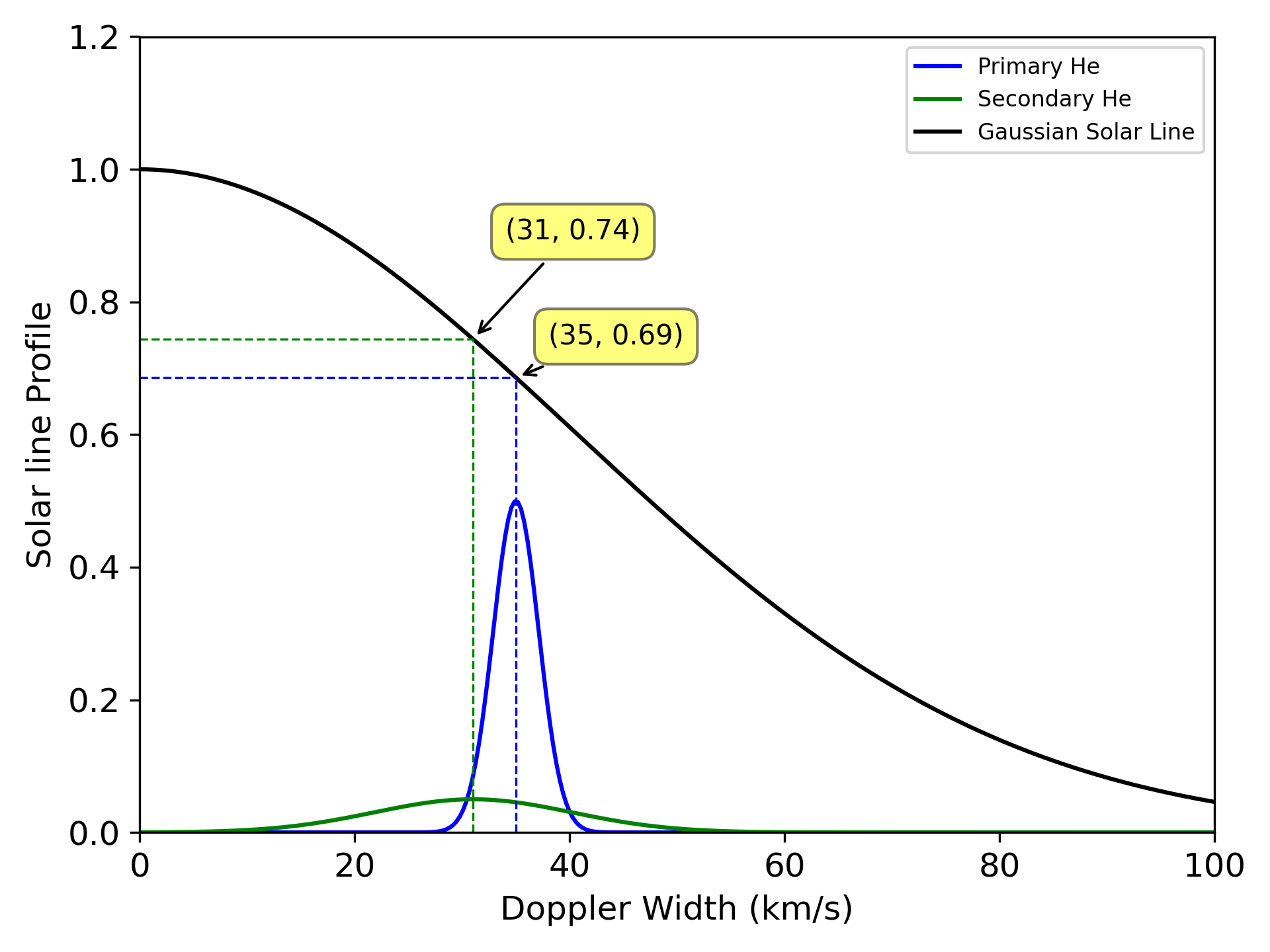}
    \caption{This figure illustrates the overestimation of the excitation factor $g$ due to the presence of secondary helium. The wide, green Gaussian distribution representing secondary helium results in a higher excitation factor compared to the much narrower, solid blue distribution of primary helium. This difference is attributed to the lower velocity of secondary helium. The radial velocities of both helium populations are calculated at 1 AU, assuming an angle of 45° from perihelion. The solar line profile, depicted by the solid black line, demonstrates the variation in solar spectral line intensity across different Doppler velocities.}
    \label{fig:euv}
  \end{figure}

Once the ISN helium parameters (Temperature and Velocity vector) influencing the intensity distribution  pattern are determined, the product \( g_0 n_{\infty} \) can be derived as the ratio of the observed intensity to the predicted intensity from a model where \( g_0 n_{\infty} = 1 \) \citep{1984A&A...134..171D}, where the excitation factor \( g_0 \) represents the rate at which a helium atom at rest scatters photons at 1 AU and $n_{\infty}$ is the density of helium in the interstellar medium. The excitation factor is defined as \( g_0 = F_{\lambda 0} \sigma_{\lambda 0} \), where \( F_{\lambda 0} \) is the solar line flux at \( \lambda = 584 \, \text{Å} \), and \( \sigma_{\lambda 0} \) is the cross section for scattering photons by helium at rest. Helium atoms moving at different speeds experience different resonance frequencies due to the Doppler shift, resulting in varying cross sections for photon scattering and consequently different excitation factors. As redistributed helium atoms from the flux core will also contribute to the resonance emission, this does not affect the density estimation. However,  due to it's asymmetry in the velocity space relative to the primary component, the presence of secondary helium slightly alters the excitation factor \( g \), leading to a minor change in the density estimation, as demonstrated in Figure \ref{fig:euv}. In the helium flux core, primary helium arrives at 1 AU at a 45-degree angle relative to the observer-sun line with a speed of approximately 35 km/s. For secondary helium, this speed is around 31 km/s. Due to the presence of secondary helium, the solar line flux \( F_{\lambda} \) is slightly overestimated, by about 7\%. Since secondary helium constitutes only about 6\% of the primary helium, the excitation factor is overestimated by approximately 0.4\%, which has a negligible effect on the density estimation. Thus, the density estimated by this method is unaltered.  Here we assume that the resonant scattering cross section is constant for both velocities.
\subsection{PUIs Observation by SWICS/Ulysses}
The density estimation by the SWICS instrument via He$^{++}$ pickup ions is the most accurate method for estimating ISN helium density for two reasons. First, the observations are made far from the Sun, at around 5 AU, where the photoionization rate is negligible. The uncertainty in calculating the photoionization rate is a fundamental source of error in density estimation. Second, no absolute geometric factor calibration is required for the density estimation. The only free parameter involved in the calculation is the cross section for PUI production, which is beyond the scope of this study. As the two populations of helium are responsible for the production of pickup ions in the inner heliosphere, no correction is needed for that purpose. Additionally, the redistributed helium is expected to generate pickup ions (PUIs) in a manner similar to primary helium. 
\subsection{Direct Detection by GAS/Ulysses} Prior to  the observational verification of secondary helium, the scientific consensus was that, unlike ISN hydrogen, ISN helium did not interact with the boundary of the heliosphere. Consequently, the density estimates derived from direct detection were considered to reflect the density in the pristine interstellar medium. However, following the discovery of the Warm Breeze by \citet{2014ApJS..213...29K}, it was determined that approximately 5.7\% of ISN helium is indeed converted into secondary ISN helium. Given that the signal-to-noise ratio for ISN helium detected by Ulysses is at best around 5, it was not possible to observe the contributions from secondary helium and the redistributed flux of primary helium. Therefore, a 5.7\% filtration correction must be included in the density estimation.\\
To incorporate scattering loss at first the location of the Ulysses spacecraft has to be determined for the relevant time period, followed by the solar wind parameters. Ulysses is the only spacecraft which orbits the sun in a plane almost perpendicular to the ecliptic. From the time of it's launch in 1990, Ulysses has made three fast latitude scans around the sun. The direct observation of helium data analyzed by \citet{2004A&A...426..835W} uses first two fast latitude scans around perihelion (1994/09-1996/08 and 2000/09-2002/08) and \citet{2015ApJ...801...62W} uses all three fast latitude scans including the last one (2006/09- 2007/08) to estimate the density of ISN helium and other parameters. We revise the density estimation from \citet{2015ApJ...801...62W} only for the final scan to avoid giving disproportionate weight to the first two scans. Note that, the photoionization rate used by these authors are significantly higher than the other studies, consequently their estimated density is also higher.
\begin{figure*}[t]
  \centering
  \begin{minipage}{0.90\textwidth}
    \centering
    \includegraphics[width=\textwidth]{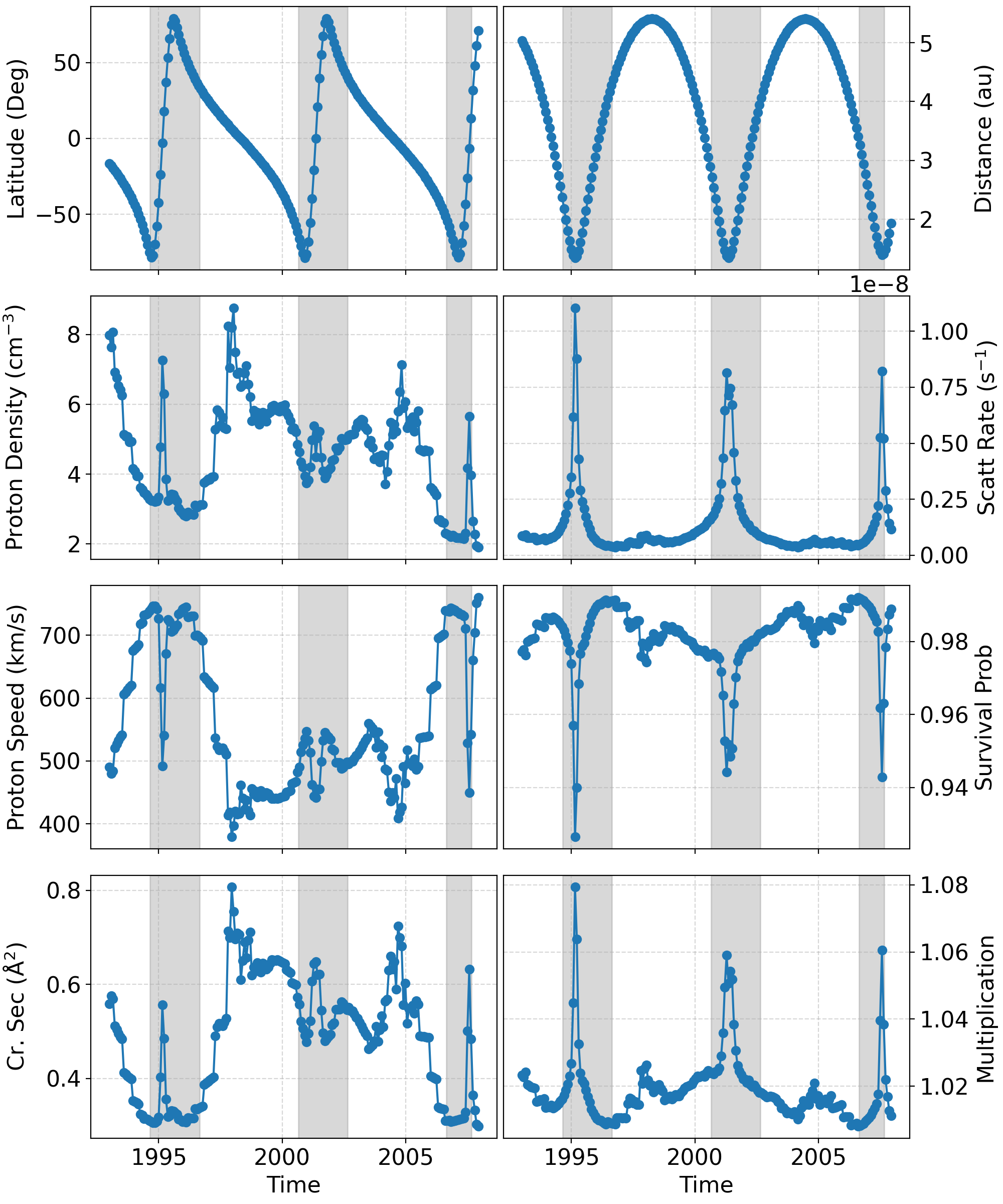}
    
  \end{minipage}
    \caption{ Position of the Ulysses spacecraft in the ecliptic coordinates , along with the different component of the solar wind and loss parameters of ISN helium. The two panels above give the latitude and heliocentric distance of Ulysses from the Sun. Gray-shaded regions represent the three periods of rapid latitude scans during which ISN helium was measured. During each of these three periods, Ulysses swept through near perihelion at high speed.Panel 2 and Panel 3 plot the solar wind proton density and velocity, respectively. The elastic scattering cross section as a function of relative velocity between protons and helium atoms is plotted in Panel 4. In Panel 6, we plot the rate of scattering loss and the corresponding survival probability in Panel 7.Panel 8 finally provides the correction factor, defined as the inverse of the survival probability. See the text below for a detailed explanation.}
    \label{fig:plot_uly}
\end{figure*}
Based on Ulysses' location, we estimate the solar wind speed and density, with the values provided in 10-degree bins ranging from 90 degrees to -90 degrees. For any specific latitude, the average is taken between the neighboring upper and lower latitude bins. For instance, if the position is at 55 degrees, we take the average between 50 and 60 degrees. The scattering cross section is calculated using equation \ref{eq:sigma}. Once the scattering cross section is determined, the scattering rate at a distance \( r \) from the Sun, the associated survival probability, and the correction factor $f_{sc}$ are subsequently calculated as described in the previous section. We compute $f_{sc}$ over the period during which the data was analyzed and determine the mean $f_{sc}$. The standard deviation of these values represents the error associated with the factor.Figure \ref{fig:plot_uly} illustrates Ulysses' position and various solar wind parameters at 1 AU from 1992 to 2008. The shaded gray regions indicate the periods of fast latitude scans. The top left panel shows Ulysses' latitude, while the subsequent bottom two panels display the solar wind proton density and speed at that latitude at 1 AU. The bottom left panel shows the elastic scattering cross section based on the velocity. From the solar wind proton density, speed, and cross section, the scattering rate is calculated, which is then adjusted according to Ulysses' heliocentric distance, as shown in the top right panel. The corresponding scattering rate at that distance is depicted in the next bottom panel. The following two panels display the survival probability of helium atoms and the correction factor, $f_{sc}$. The sharp increase of $f_{sc}$ during the first three fast latitude scans is primarily due to the Ulysses passing through perihelion. \\
The newly revised density estimates from GAS/Ulysses, along with those from two other methods, are presented in Figure \ref{fig:all_density}. The x-axis represents the distance from the Sun. The density estimated from resonant scattering of He 584Å is at 1 AU, while the pickup ion method provides estimates at 5 AU. The GAS/Ulysses density covers a range from 1.3 AU to 2.5 AU, but for clarity, these values are plotted near 2.5 AU. Previously estimated densities are marked by filled circles, and the revised estimates by filled diamonds. \citet{2004A&A...426..897M} reported a combined density of 0.0148 $\pm$ 0.002 from all three methods, reflecting the consolidated local interstellar medium parameters. After revising the GAS/Ulysses data, the updated combined density is 0.0153 $\pm$ 0.0011, which remains within the upper bound of the previously estimated combined density. The original and revised combined densities are depicted by red filled circles and diamonds, respectively.

In SC 24, the average scattering rate of ISN helium atoms by solar wind protons, which redistributes atoms from the flux core, is approximately $0.15\times 10^{-7} \text{s}^{-1}$. This results in an additional reduction in survival probability by 10\%, which is a non-negligible effect. This additional loss implies that any estimation of ISN helium density via direct detection must account for redistribution loss.

%% file: sections/7conclusion.tex
\section{Discussion and Conclusion}\label{sec:conclusion}

We analyzed the observations  of primary and secondary interstellar neutral helium with IBEX-Lo and examined  the temporal variation of $\textbf{A}_{\text{p}}$ and $\textbf{A}_{\text{s}}$ from 2013 to 2020 and found the same trend for primary helium as reported by \citep{2023ApJ...953..107S}, which is a gradual increase of $\textbf{A}_{\text{p}}$ from 2015 onwards. However, a systematic decrease in \( \textbf{A}_{\text{s}} \) from the solar maximum is observed when \( \textbf{A}_{\text{p}} \) is used (Figure \ref{fig:all_images}). If \( \textbf{A}_{\text{p}} \) is not included in the calculation of \( \textbf{A}_{\text{s}} \), the trend reverses, becoming similar to the trend of \( \textbf{A}_{\text{p}} \). For both of the cases a time independent average value of \( \textbf{A}_{\text{s}} \) is ~ 20\% suggesting contribution from additional source. \\
We established that the scattering of interstellar helium by solar wind protons creates a halo, which naturally explains the increased flux of secondary helium. A higher correlation coefficient, when \( \textbf{A}_\text{p} \) is excluded from the calculation of \( \textbf{A}_\text{s} \), suggests that the primary helium population near the peak is not modulated in the same way as it is away from the peak.\\
We demonstrated that the density estimated from the direct detection of ISN helium by GAS/Ulysses needs to be revised to account loss by scattering  as well as for the filtration of primary helium in the OHS, which contributes to the creation of the Warm Breeze.
The scattering angle of ISN helium by solar wind protons depends on their relative speed; at higher speed, both the scattering cross section and scattering angle decreases, leading to very minor deviation from the original trajectory. The limitations of our analysis include the lack of velocity dependent scattering angle.Also, We assume that only 10\% of the elastic scattering interaction produces an observable halo, and we calculate the scattering rate using the solar wind flux in the ecliptic plane at 1 AU. To enhance the robustness of this analysis, a comprehensive 3D model that tracks particle trajectories and incorporates the appropriate scattering rate is required.\\
In conclusion, the only methodology that requires revision for density estimation is the direct detection by the Ulysses GAS experiment.The revised combined density, based on earlier estimates, is calculated to be \(0.0153 \pm 0.0011\). The original and revised values are provided in Table \ref{tab:result}.
\begin{figure*}[t]
  \centering
  \begin{minipage}{0.75\textwidth}
    \centering
    \includegraphics[width=\textwidth]{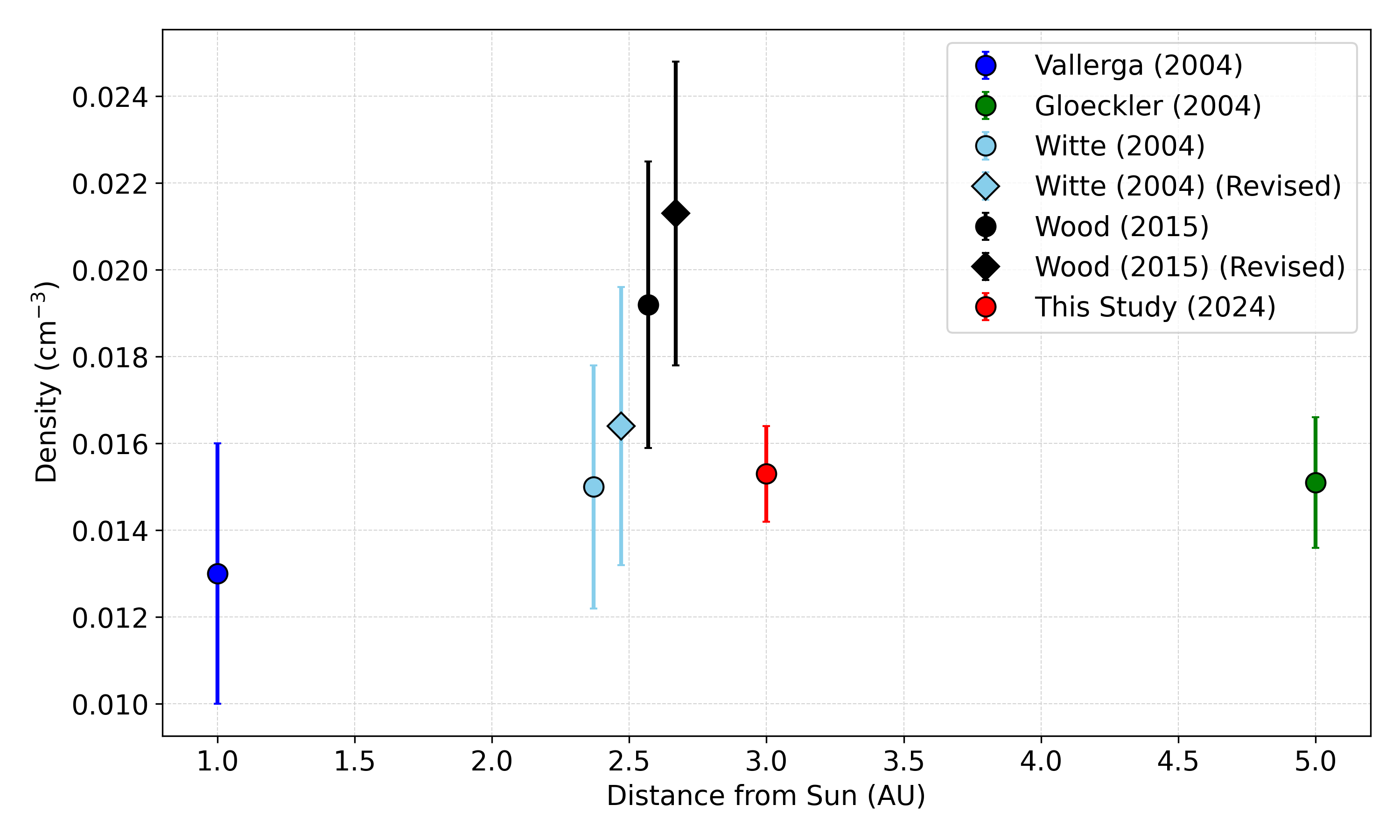}
    \caption{Reassessment of the previously estimated density by various studies. The density determined by \citet{2004A&A...426..855V}  and \citet{2004A&A...426..845G}  remain unaltered. Revisions to the Ulysses/GAS estimations account for the influence of the warm breeze and scattering effects. The combined weighted average density estimated in this study is denoted by the red filled circle.}
    \label{fig:all_density}
  \end{minipage}
\end{figure*}
Since the absolute geometric factor for IBEX was not determined with sufficient precision, accurate density estimation of ISN helium or hydrogen from IBEX-Lo observations is not feasible. However, future instruments like IMAP-Lo should be able to account for the scattering processes in more detail. If the loss in the helium flux core is considered, the core flux observed at 1 AU decreases by approximately 10\%.

We thank all individuals associated with IBEX. We specially thank Paweł Swaczyna for many helpful discussions. This work is supported by the Interstellar Boundary Explorer mission as part of NASA’s Explorer Program and partially by NASA SR\&T Grant NNG06GD55G. JMS has been supported by NASA grant 80NSSC20K0719.